\documentclass{article}

\usepackage{PRIMEarxiv}

\usepackage[utf8]{inputenc} % allow utf-8 input
\usepackage[T1]{fontenc}    % use 8-bit T1 fonts
\usepackage{hyperref}       % hyperlinks
\usepackage{url}            % simple URL typesetting
\usepackage{booktabs}       % professional-quality tables
\usepackage{amsfonts}       % blackboard math symbols
\usepackage{nicefrac}       % compact symbols for 1/2, etc.
\usepackage{microtype}      % microtypography
\usepackage{lipsum}
\usepackage{fancyhdr}       % header
\usepackage{graphicx}       % graphics
\usepackage{amsmath} 
\usepackage{amssymb}
\usepackage{algorithm}
\usepackage{algpseudocode}
\usepackage{multirow} 
\usepackage{pifont}
\usepackage{enumerate}
\graphicspath{{media/}}     % organize your images and other figures under media/ folder

\usepackage{xcolor}
\usepackage{soul}
\usepackage{verbatim}

\title{WaveDiffUR: A diffusion SDE-based solver for ultra magnification super-resolution in remote sensing images 
}

\author{
  Yue Shi, Liangxiu Han*, Darren Dancy \\
  Department of Computing and Mathematics, \\
  Manchester Metropolitan University \\
  Manchester, UK\\
  \texttt{y.shi@mmu.ac.uk, l.han@mmu.ac.uk} \\
   \And
  Lianghao Han \\
  Department of Computer Science \\
  Brunel University \\
  London, UK\\
  \texttt{} \\
}

\begin{document}
\maketitle

\begin{abstract}
Deep neural networks have recently achieved significant advancements in remote sensing superresolution (SR). However, most existing methods are limited to low magnification rates (e.g., $\times 2$ or $\times 4$) due to the escalating ill-posedness at higher magnification scales. To tackle this challenge, we redefine high-magnification SR as the ultra-resolution (UR) problem, reframing it as solving a conditional diffusion stochastic differential equation (SDE). In this context, we propose WaveDiffUR, a novel wavelet-domain diffusion UR solver that decomposes the UR process into sequential sub-processes addressing conditional wavelet components. WaveDiffUR iteratively reconstructs low-frequency wavelet details (ensuring global consistency) and high-frequency components (enhancing local fidelity) by incorporating pre-trained SR models as plug-and-play modules. This modularity mitigates the ill-posedness of the SDE and ensures scalability across diverse applications.
To address limitations in fixed boundary conditions at extreme magnifications, we introduce the cross-scale pyramid (CSP) constraint, a dynamic and adaptive framework that guides WaveDiffUR in generating fine-grained wavelet details, ensuring consistent and high-fidelity outputs even at extreme magnification rates. Extensive experiments demonstrate that the WaveDiffUR model, combined with CSP, achieves state-of-the-art performance by effectively minimizing degradation across key metrics: quantitative accuracy, perceptual quality, spectral consistency, and sharpness. Even when the magnification scale increases significantly from $\times 4$ to $\times 128$, the model maintains robust performance, with an average degradation of only $19.1\%$. At extreme magnifications (e.g., $\times 128$), it outperforms benchmark models, achieving up to 3 times the improvement in PSNR and SRE, showcasing superior image quality and spectral fidelity. By enabling robust ultra-resolution in remote sensing, WaveDiffUR opens new possibilities for applications in environmental monitoring, urban planning, disaster response, and precision agriculture. This study provides a significant step toward scalable, cost-effective, and high-fidelity solutions for real-world remote sensing challenges.
 
\end{abstract}
%-- **** in the abstract, please provide experiment results with the improvements figures, for instance, improve by how many percentage xxxxxx, compared to xxxxx methods **

% keywords can be removed

\section{Introduction}

Remote sensing image super-resolution (SR) remains a persistent challenge and continues to be a vibrant research topic in both computer vision \cite{wang2022comprehensive} and geosciences \cite{wang2022review}. SR aims to reconstruct high-resolution (HR) remote sensing observations with realistic spectral-spatial details from low-resolution (LR) data, which are typically acquired from aerial platforms (1 - 10m resolution) or space platforms (>10 m resolution). However, current SR research predominantly focuses on fixed and low-magnification scales (e.g., $\times 2$ or $\times 4$) \cite{zhang2023superyolo, xiao2023ediffsr, an2023efficient}, which fails to meet the high-magnification scale SR demands for various Earth observation tasks. For example, land-cover mapping typically requires a spatial resolution of 1-2 meters, necessitating $\times 8$ SR for 10-meter Sentinel-2 data or $\times 16$ SR for 30-meter Landsat-8 data \cite{he2023very}. Precision agriculture demands resolutions higher than 1 meter, translating to $\times 16$ SR for Sentinel-2 data or $\times 32$ SR for Landsat-8 data \cite{shi2022novel}. Additionally, multi-modal data fusion requires resolution matching between different data sources, which requires various scaling-rate SR enhancements ranging from $\times 4$ to $\times 64$ \cite{he2023self}.\par

\begin{figure}[]   
    \centering  
    \includegraphics[width=5in]{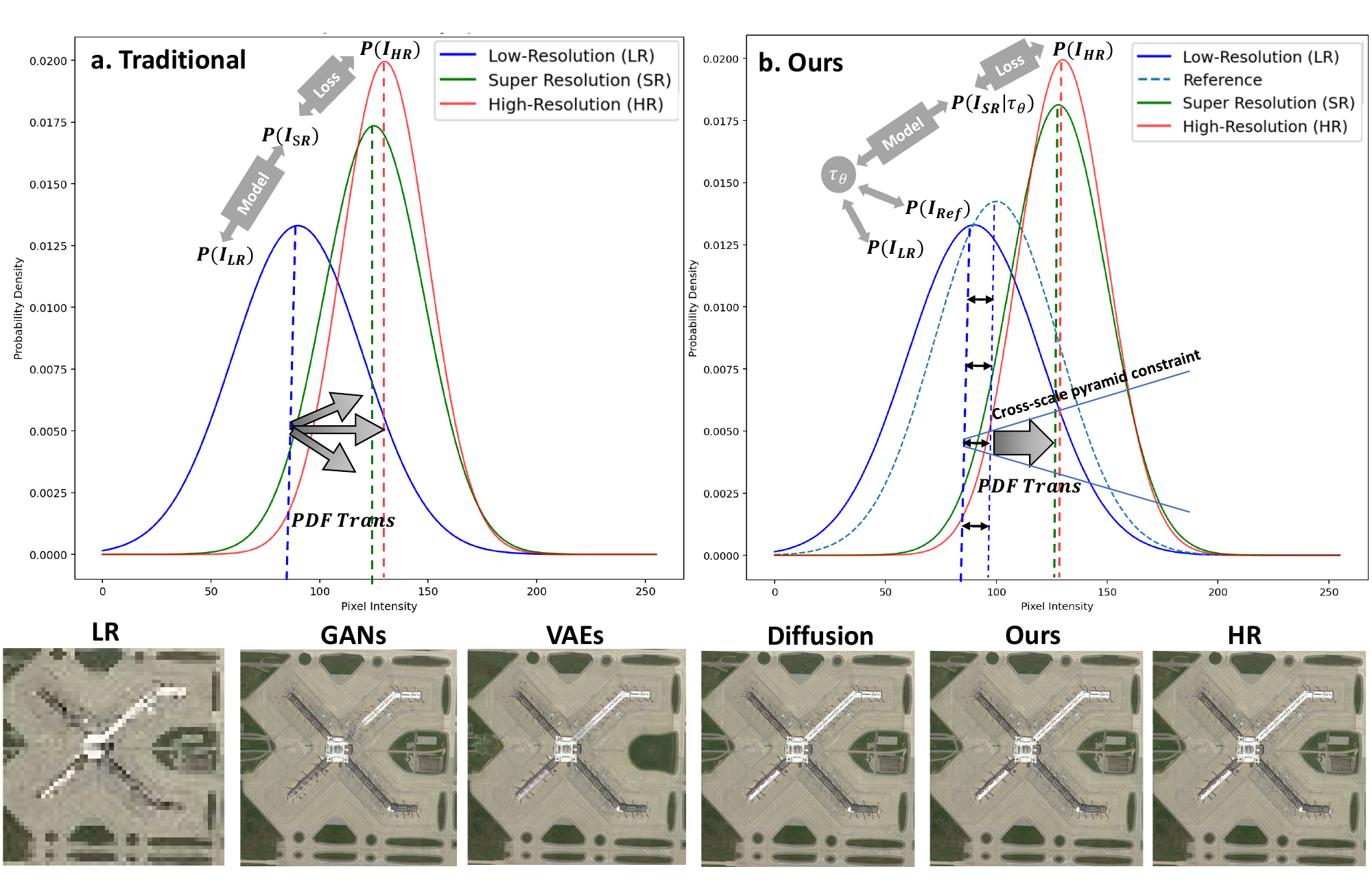}  
    \caption{A comparison of SR performance across several existing PDF-based SR models using deep learning approaches, including our previous model \cite{shi2022latent}. }
    \label{fig:0_0}  
\end{figure}

Unlike human or natural image super-resolution (SR), remote sensing image SR faces challenges unique to Earth Observation (EO), such as complex spatial heterogeneity and the presence of mixed pixels. These factors exacerbate the inherently ill-posed nature of SR at high-magnification scales, where the HR counterpart may exist in an infinite solution space for a given LR input \cite{fu2024continuous}. Existing deep learning-based SR methods primarily address these challenges by training neural networks to model the probability density function (PDF) transition in the pixel-wise representation space, mapping the PDF of LR images to that of their HR counterparts. Fig \ref{fig:0_0} illustrates a comparison of SR performance across the most popular PDF-based SR models using deep learning approaches, including our previous model \cite{shi2022latent}. Among them, generative adversarial networks (GANs) based models use adversarial learning between a generator and a discriminator to produce SR images with realistic HR details missing in the LR images, to align the SR PDF more closely with the target HR PDF \cite{zhu2020gan, song2023esrgan}. Variational autoencoders (VAEs), on the other hand, encode the LR PDF into a latent space and then generate the SR image though sampling it, aiming to match the reconstructed SR PDF to that of the HR image \cite{liu2020photo}. However, due to the ill-posed nature of SR, the current models are predominantly trained on LR-HR image pairs with low magnification rates (e.g., $\times 2$ or $\times 4$), which provide effective cross-scale representations to address the problem. Our previous study \cite{shi2022latent} attempted to explore a GAN-based approach at high magnification rates. It demonstrated that when the magnification scale exceeds $\times 8$, the SR quality and fidelity diminish due to issues of mode collapse and perceptual artifacts. This decline arises from the adversarial nature of the GAN-based model, which is notoriously difficult to achieve convergence given the increased complexity of PDF transition in very large-magnification SR tasks \cite{zhu2020gan}. At high magnification rates, the discontinuity of the cross-scale representations will hinder the model's ability to maximize the likelihood of SR outputs given LR inputs. This makes the transition from the LR PDF to the HR PDF significantly more complex and less predictable. \par

Recently, diffusion models (DMs) \cite{ho2020denoising} have received increasing attention in the field of image restoration and also demonstrated promising performance in remote sensing image SR \cite{ xiao2023ediffsr, an2023efficient, liu2022diffusion}. The strength of diffusion models (DMs) lies in their denoising diffusion process, which gradually transforms the LR PDF into the HR PDF through a series of small, incremental noise removal steps. This process enables DMs to model high-quality SR data distributions while addressing common issues of training instability and mode collapse  observed in GAN-based models, thanks to their well-defined probabilistic framework. Despite these advantages, remote sensing image UR presents a unique challenge, as it requires generating realistic and consistent spectral-spatial information across large-scale spans. Diffusion models, which are inherently based on stochastic processes, may face difficulties in consistently inferring and reconstructing the fine spectral-spatial details over extensive areas. \par

In this study, we introduce the concept of a cross-scale pytimid (CSP) boundary condition, which represents the unmixing rules of spectral-spatial scaling details across different magnification levels. Building on this concept, we formulate the UR process as a conditional diffusion SDE for low-frequency (fidelity) information upscaling and high-frequency (consistency) detail restoration. To solve this equation, we propose a wavelet-based diffusion UR SDE solver, named WaveDiffUR, which guides the conditional inverse diffusion process in the wavelet domain to alleviate the ill-posed nature of UR. The WaveDiffUR framework, as shown in Fig \ref{fig:1w}, enables the seamless integration of pre-trained SR pipelines as a plug-and-play module to generate the cross-scale condition, significantly reducing the costs associated with training new models from scratch. However, re-utilizing a fixed and unique condition throughout the progressive UR process may degrade the constraint capacity for solving the inverse diffusion SDE, thereby compromising the consistency and fidelity of the UR results. To address this limitation, we introduce a dynamically updated cross-scale condition, named cross-scale pyramid (CSP). The CSP serves as a variable boundary condition for the SDE solver by compressing information from adjacent UR sub-processes, guiding the WaveDiffUR solver to generate accurate UR results with realistic spectral-spatial details. Experimental results demonstrate that the baseline WaveDiffUR model without CSP exhibits high performance in terms of usability, adaptability, and cost-effectiveness. Moreover, the enhanced CSP-WaveDiffUR model effectively captures the unmixing rules of realistic spectral-spatial details, thereby improving UR efficiency and robustness in handling high-magnification SR tasks. \par

The primary contributions of this work are as follows: \par
\begin{enumerate}[(i)]
        \item Pioneering the Utra-Resolution (UR)  Problem: To the best of our knowledge, this is the first work to explicitly address the complex ill-posed UR problem. We develop a universal SDE solver, termed WaveDiffUR, which decomposes the complicated UR process into  finite sub-processes. It maximizes the utilization of pre-trained SR models, enhancing the usability, adaptability, and cost-effectiveness of the UR process.
        \item Dynamic Boundary Condition for the SDE solver (CSP-WaveDiffUR): To address the degradation issue caused by re-using fixed and unique boundary conditions in constraining the diffusion UR SDE, we propose an improved version of the SDE solver, CSP-WaveDiffUR. This model dynamically updates boundary conditions during each UR sub-process, ensuring high-quality UR results with improved consistency and fidelity.
        \item Extensive Experimental Validation: Through comprehensive experiments, we validate the effectiveness of both the universal WaveDiffUR and the enhanced CSP-WaveDiffUR models and their ability to maintain UR consistency and fidelity in high-magnification UR tasks.
    \end{enumerate}

By addressing fundamental challenges in ultra-resolution, WaveDiffUR opens new opportunities for practical remote sensing applications, including environmental monitoring, urban planning, disaster response, and precision agriculture. This study presents a scalable, cost-effective, and high-fidelity approach to advancing remote sensing capabilities at unprecedented magnification levels. \par

The remainder of this paper is organized as follows: Section \ref{sec:rw} review the related work on diffusion model and hyperspectral image super-resolution. Methodology is detailed in Section \ref{sec:method}, including the main framework of the proposed WaveDiffUR method. Experimental results are reported in Section \ref{sec:ed}. Finally, Conclusion are presented in Section \ref{sec:con}.\par

\section{Related work}
\label{sec:rw}
\subsection{Remote sensing image super-resolution}
To transform low-resolution remote sensing images into high-resolution counterparts, considerable efforts have been made to improve image fidelity and restoring details. Traditional methods mainly adopt fusion technologies, such as wavelet transform and spectral mixing analysis, in which the spatial information of high-resolution images can be used to improve the spatial details of low-resolution images effectively. For instance, Zhang \textit{et al} \cite{zhang2007multi} introduced an image fusion technique utilizing the 3D Wavelet Transform (3DWT). This method is particularly effective for remote sensing imaging, where the spectral dimension information is crucial, unlike in panchromatic or RGB images. The 3DWT excels in harnessing this spectral information to produce higher-quality fused images. Yokoya, \textit{et al.} \cite{yokoya2011coupled} proposed a coupled non-negative matrix factorization (CNMF) algorithm, based on a linear spectral mixing model. This algorithm reconstructs images with both high spatial and spectral resolution by merging low spatial resolution hyperspectral images (HSI) and high spatial resolution multispectral images (MSI), incorporating their structural features. Furthermore, realising the potential of frequency-domain gain recognition in the field of Super-Resolution (SR), Zhang \textit{et al} \cite{zhang2009noise} proposed a Bayesian estimation approach for multispectral images in the wavelet domain. This approach demonstrates strong noise resistance and consistently produces reliable fusion results, showcasing its effectiveness in handling spectral-spatial information in SR applications. \par

In recent years, significant advancements have been made in deep learning-based approaches to the SR problem. Dong \textit{et al}. \cite{dong2014learning} first introduced Super-Resolution Convolutional Neural Network (SRCNN) model to learn the mapping relationship between low-resolution images (LR) and the corresponding high-resolution (HR) images, in which the LR images were produced by upscaling original LR images to the desired size using bicubic interpolation. Since then, several innovative convolutional neural network (CNN) models have been developed, emphasizing the critical role of network architecture design in enhancing image reconstruction performance of SR. For example, Muhammad, \textit{et al.} \cite{muhammad2023senext} implemented a "Squeeze-and-Excitation" module to incorporate an attention mechanism targeting the channel dimension. This method utilizes a compact network designed to autonomously ascertain the significance of each channel, and then allocates network weights to each feature accordingly, enabling the network to focus on the most relevant feature channels. Similarly, Zheng, \textit{et al.} \cite{zheng2020hyperspectral} applied spatial-spectral attention mechanism to neural network models for panchromatic sharpening of hyperspectral images (HSI), allowing the networks to adaptively learn both spatial and spectral details.

In general, deep learning-based approaches in improving spatial resolution of remote sensing imaging primarily adopt two strategies: fusion with other high spatial resolution images \cite{zheng2023msisr, han2019multi, chen2021remote} and single-image based SR \cite{chen2022real, behjati2023single}. Fusion-based SR techniques utilize additional external prior information to reconstruct images with finer textures. In contrast, single-image-based SR techniques do not rely on any auxiliary data, offering greater practical feasibility. For instance, Mei \textit{et al.} \cite{mei2017hyperspectral} developed a 3D Fully Convolutional Neural Network (3D-FCNN) model for drone image super-resolution, incorporating an upsampling process at the earlier stage. Jiang \textit{et al.} \cite{jiang2020learning} proposed the Single Sub-Image Progressive Super-Resolution (SSPSR) model, which involves progressive sampling: first for grouped sub-images and then for the entire image constructed by fusing the interpolated sub-images. This approach improves feature extraction in HSIs and enhances the stability of overall training, although it will also introduce extra requirements, such as more precise modeling and intricate network design for each stage.\par

The Generative Adversarial Network (GAN) is another deep learning based model that has gained significant attention in the field of super-resolution. GANs are particularly valued for their ability to model complex data distributions, enabling the generation of high-resolution (HR) images that closely mimic the quality and perceptual characteristics of real-world data. When integrated into the SR process, GANs generate HR images with enhanced visual appeal. Xiong \textit{et al}. \cite{xiong2020improved} proposed an improved Super-Resolution Generative Adversarial Network (SRGAN) with a revised loss function and optimised network architecture. These modifications improve the training stability and generalisation performance. Shi \textit{et al} \cite{shi2022latent} proposed the latent encoder integrated GAN, named LE-GAN, incorporating self-attention mechanisms to boost feature extraction of the generator and stabilize the training process. \par

In parallel, diffusion probabilistic models (DPMs) have emerged as another promising approach for super-resolution tasks. DPMs generate high-quality data distributions through a structured and well-defined probabilistic diffusion process, mitigating the trainning instability often seen in GANs. Recently, Saharia et al. \textit{et al.} \cite{saharia2022image} advanced a DPM-based super-resolution method, employing a UNet architecture as the denoiser to iteratively refine image generation. Luo \textit{et al.} \cite{luo2023image} further enhanced diffusion-based SR by introducing stochastic differential equations (SDEs) to better model the degradation of diffusion process. These developments underscore the potential of diffusion models for addressing complex SR challenges. A detailed investigation on diffusion-based super resolution are presented in the following section. \par

\subsection{Diffusion-based image super-resolution}
Diffusion-based models, especially denoising diffusion probabilistic models (DDPMs) \cite{ho2020denoising}, have emerged as a promising approach for image restoration tasks such as super-resolution \cite{xu2023dual, li2022srdiff, xiao2023ediffsr}, inpainting \cite{zhu2023denoising, corneanu2024latentpaint}, and deblurring \cite{spetlik2024single, whang2022deblurring}. For example, Kawar et al. \cite{kawar2022denoising} introduced Denoising Diffusion Restoration Models (DDRM), which utilizes a pre-trained diffusion model to solve various linear inverse problems, showing superior performance across multiple image restoration tasks. Wang \textit{et al.} \cite{wang2023dr2} proposed DR2, Diffusion-Based robust degradation remover for  Blind Face Restoration, which initially uses a pre-trained diffusion model for coarse degradation removal then followed by an enhancement module aiming for finer blind face restoration. Guo \textit{et al.} \cite{guo2023shadowdiffusion} developed ShadowDiffusion, which leverages an unrolled diffusion model to address the challenging task of shadow removal by progressively refining results with degradation and generative priors. \par

The basic principle of diffusion-based image super-resolution involves the use of a Markov chain to model the transformation of high-resolution (HR) image data into noise and back again \cite{kawar2022denoising}. It is composed of two opposite processes: Forward process (Diffusion process) and Reverse Process (Denoising with condition). The forward process gradually corrupts an HR image through a Markov chain and transitions the HR image distribution into a stochastic (Gaussian noise) distribution by progressively adding noise. The process effectively creates a dataset of noisy images that represent the HR data in a stochastic space. In the reverse process, the HR image is reconstructed from the noisy data using the corresponding low-resolution image as a conditional factor to guide the systematic removal of noise. The model transforms the noisy data back into the HR distribution by iteratively refining the image. Through this denoising process, the conditional diffusion model ensure the spatial and spectral consistency with the LR input while reconstructing fine details of the noisy data. \par 

The forward noising process of diffusion-based image super-resolution can be defined with the following stochastic differential equation (SDE)\cite{song2020score}: 

% \begin{comment}{The principle of diffusion-based image super-resolution is to train a Markov chains which entails the transformation of a high-resolution image dataset, denoted as $I_{HR} \in \mathbb{R}^{n \cdot H \times n \cdot W \times C}$, into entirely stochastic data by incrementally introducing noise. Subsequently, the corresponding low-resolution image dataset $I_{LR} \in \mathbb{R}^{H \times W \times C}$ serves as a conditional factor to facilitate the systematic prediction and elimination of this noise, thereby generate the super-resolution image dataset $I_{SR} \in \mathbb{R}^{n \cdot H \times n \cdot W \times C}$. By sampling a HR-LR image pair (i.e. $x \in I_{HR}$ and $x \in I_{LR}$), the forward noising process can be defined with the following stochastic differential equation (SDE)\cite{song2020score}:
% \end{comment}

\begin{equation}
\label{fuc:1}
dx = \bar{f}(x, t) dt + \bar{g}(t) dw
\end{equation}

where $\bar{f}(x, t)$ is the linear drift function which governs the rate at which the HR image data $x$ is noised in time-step $t$,  $\bar{g}(t)$ is a scalar diffusion coefficient associated with $t$, and $w$ denotes the standard Wiener process. \par

% \begin{comment}
% \begin{equation}
% \label{fuc:2}
% I_{SR} = H \cdot I_{LR} + \epsilon_{hf}
% \end{equation}

% where $H$ serves as the cross-scale pyramid matrix which indicates the mapping rule between the $I_{SR}$ and $I_{LR}$. $\epsilon_{hf}$ are the high-frequency components that restore the spectral-spatial details towards the SR image. Therefore, according to Anderson’s theorem \cite{chung2022improving, song2020score}, the reverse diffusion process can be represented by a reverse SDE function:
% \end{comment}

Using Anderson’s theorem \cite{chung2022improving, song2020score}, the reverse diffusion process in SR can be represented by a reverse SDE function :
\begin{equation}
\label{fuc:3}
dx = \left[\bar{f}(x, t) - \bar{g}(t)^2 \nabla_x \log p_t(x \mid y)\right] dt + \bar{g}(t) d\bar{w}
\end{equation}

where $x$ represents the reconstructed HR image, $y$ acts as a condition that is typically derived from the LR image or intermediate features, $dt$ denotes the infinitesimal negative time step, and $\bar{w}$ defines the reverse Wiener process. The reverse SDE defines the generative process through the score function $\nabla_x \log p_t(x \mid y)$ and minimizes the following denoising score-matching objective:

\begin{equation}
\label{fuc:4}
\min_{\theta} \mathbb{E}_{t \sim U(\epsilon, 1), x_0 \sim p_0(x \mid y), x_t \sim p_{0,t}(x_t \mid x_0)} \left\| s_{\theta}(x_t, t) - \nabla_{x_t} \log p_{0, t}(x_t \mid x_0) \right\|^2_2
\end{equation}

Once the parameter $\theta$ for score function is estimated, one can replace the score function $\nabla_x \log p_t(x \mid y)$ in Eq. \ref{fuc:3} with $s_{\theta}(x_t, t)$ to solve the reverse SDE. 

\subsection{Diffusion process in wavelet domain}
\label{sec:2.3}

The wavelet transformation is highly effective in reducing spatial dimensions while retaining critical information, setting it apart from transformation techniques like Fast Fourier Transformation (FFT) and Discrete Cosine Transform (DCT), which can suffer information loss during transformations. This advantage has led to its adoption in diffusion-based super-resolution (SR). The studies like Jiang et al \cite{jiang2023low}, show the benefits of performing diffusion operations in the wavelet domain rather than directly in the image space. This approach supports improved content reconstruction and narrows the disparity between ground-truth HR and SR domains. By combining wavelet transformation with diffusion model techniques, these methods achieve enhanced super-resolution outcomes and improved performance metrics. \par 
In 2D applications,  a 2D low-resolution image, $x \in I_{LR}$,  can be decomposed into four sub-bands through the 2D discrete wavelet transformation (2D-DWT) with Haar wavelet functions as follows:
\begin{equation}
\label{fuc:3.1}
\{A_{LR}, V_{LR}, H_{LR}, D_{LR}\} = \text{2D-DWT}(I_{LR})
\end{equation}

where $A_{LR} \in \mathbb{R}^{\frac{H}{2} \times \frac{W}{2} \times c}$ is the low-frequency of the input image and  $V_{LR}, H_{LR}, D_{LR} \in \mathbb{R}^{\frac{H}{2} \times \frac{W}{2} \times c}$ are respectively the vertical, horizontal, and diagonal high-frequency information of the input. Specifically, the low-frequency coefficient $A_{LR}$ encapsulates the global consistency information of the original image, acting as a downsampled representation, while the high-frequency coefficients, $V_{LR}, H_{LR} and D_{LR}$, capture sparse local details which represent the fidelity details of the image. In the field of image super-resolution, images reconstructed through upscaling high-frequency coefficients largely restore the fidelity of the original images. In contrast, the low-frequency coefficient alters the global information significantly, leading to the greatest consistency with the original images. Consequently, the main objective in image super-resolution in the wavelet domain involves retrieving the low-frequency coefficient of an LR image that aligns with its high-resolution counterpart. \par

% Although the spatial dimensions of the wavelet sub-bands are reduced to one-fourth that of the input image following a single 2D-DWT, the low-frequency information is well restored in the wavelet domain. Therefore, we proceed to apply $K$ times wavelet transformations to reduces the spatial dimension for an efficient diffusion process, specifically,
% \begin{equation}
% \label{fuc:3.2}
% \{A_L^k, V_L^k, H_L^k, D_L^k\} = \text{2D-DWT}(I_{L})
% \end{equation}

% where $A_L^k, V_L^k, H_L^k, D_L^k \in \mathbb{R}^{\frac{H}{2^k} \times \frac{H}{2^k} \times c}$, $k \in [1, K]$. Subsequently, conditional diffusion operations are conducted on the low-frequency sub-band $A_L^k$ in the wavelet domain. This process aims to efficiently enhance the resolution of low-frequency components while maintaining the 'consistency' of the global information. Correspondingly, the high-frequency sub-bands $\{V_L^k, H_L^k, D_L^k\}$ are upscaled by the proposed CSRU. In this way, our approach significantly reduces inference time and computational resource consumption of the diffusion process due to a time reduction in the spatial dimensions.

\section{Methodology}
\label{sec:method}

\begin{figure}[]   
    \centering  
    \includegraphics[width=5in]{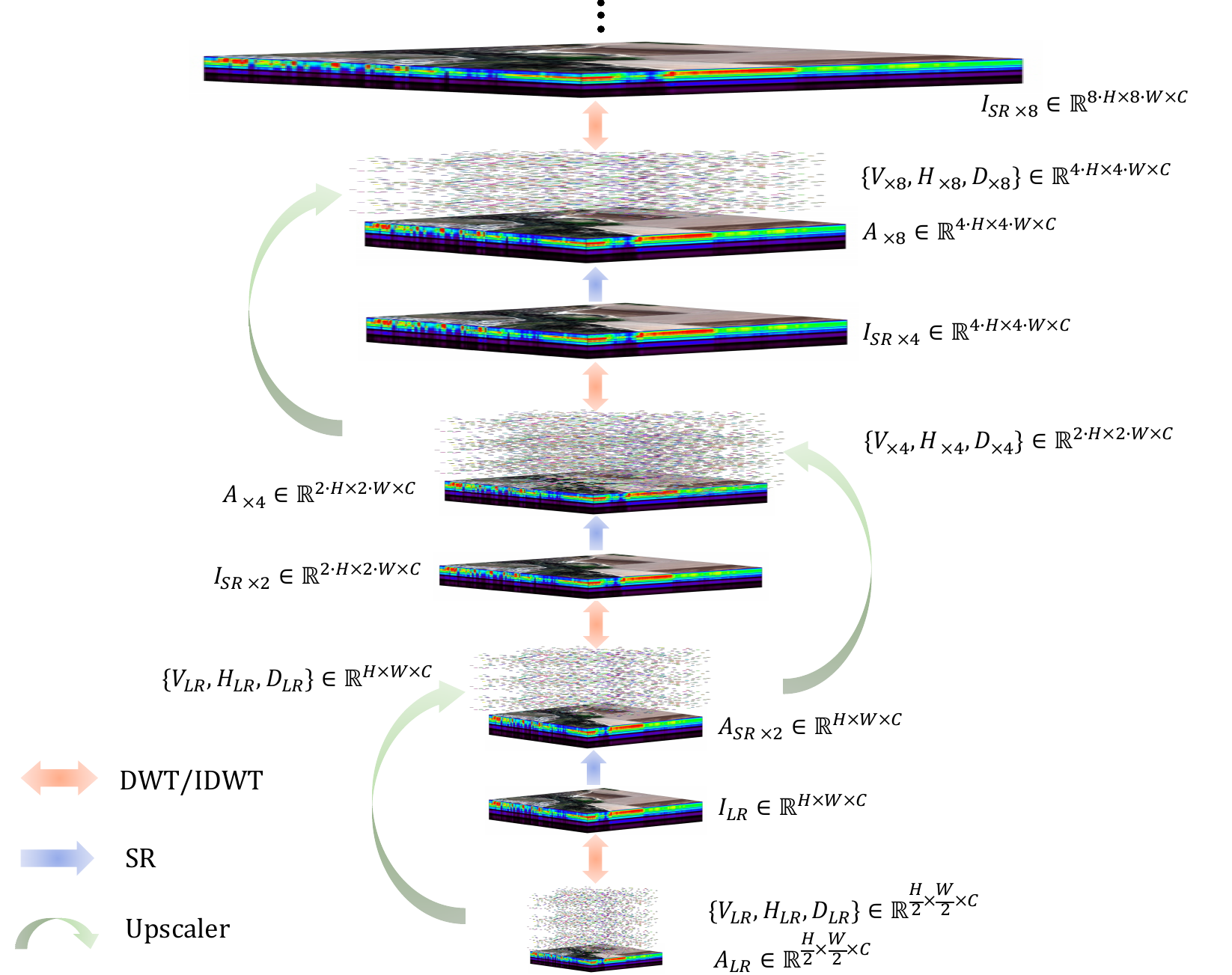}  
    \caption{An Illustration of the proposed self-cascade UR pyramid framework, consisting of 1) DWT/IDWT: cyclically decompose a low-resolution image into wavelet domain, and restore the high-resolution image from the up-scaled wavelet-domain components. 2) SR pipeline: integrates a plug-and-play tunable SR module into the framework to reconstruct the low-frequency wavelet components of a high-resolution image from it's low-resolution counterpart. 3) Upscaler: progressively adapts the high-frequency wavelet components of the low-resolution images to match those of the higher-resolution images. }
    \label{fig:1w}  
\end{figure}

We hypothesize that the ill-posed UR process can be modeled as a stochastic diffusion process of spectral-spatial unmixing, where the interplay between spectral fidelity and spatial consistency is systematically addressed. Our primary goal is to solve the ill-posed UR problem through a diffusion UR SDE and generate high fidelity and spatial consistent UR images from their low-resolution counterparts. The proposed UR process is generalized to a finite number of SR steps by formulating them as a solution to a conditional diffusion SDE, constrained by a pyramid-shaped multi-scale spectral-spatial unmixing rule. Inspired by Jiang \textit{et al} \cite{jiang2023low}, we adopt a wavelet domain implementation  for the proposed UR process to facilitate computation efficiency. The workflow of the proposed UR approach is shown in Fig. \ref{fig:1w}. The subsequent sections will elaborate on the details of the proposed method.

% Subsequently, focus on the degradation problem of the constraint condition that are prevalent in the self-cascading approach, we develop a novel SR pipeline to mitigate the computational efficiency challenges and coherence generation issues.

\subsection{Wavelet-domain diffusion UR (WaveDiffUR) SDE solver}
\label{sec:3.22}
The image UR process in the wavelet domain operates as a stochastic diffusion process of spectral-spatial unmixing from low-resolution space to high-resolution space. This is achieved through a series of SR steps formulated as solutions to conditional diffusion SDEs. Our proposed WaveDiffUR solver solves the conditional diffusion UR SDEs to enhance both LF and HF details. The solver ensures that spectral-spatial consistency is maintained, mitigates the ill-posed nature of the UR task by dynamically adapting to multi-scale constraints across magnification levels, and preserves the integrity of fine details during high-magnification upscaling. Each SR step iteratively refine the wavelet components, progressively generating UR wavelet components for each frequency band. LF components benefit from pre-trained SR modules to improve global structures, while HF components are upscaled to enhance texture and detail.  After processing the wavelet components, Inverse Discrete Wavelet Transform (IDWT) recombines them into the spatial domain. This final reconstruction step ensures that the synthesized HR image aligns with the desired fidelity and consistency of HR ground-truth data. \par

% The proposed UR SDE consist of three modules: DWT/IDWT, low-frequency wavelet feature SR, and high-frequency wavelet feature restoration. The DWT/IDWT cyclically decomposes the LR image into the wavelet-domain component and subsequently restores the UR image from the up-scaled wavelet-domain components.

Low frequency and high frequency wavelet components after DWT in WaveDiffUR can be calculated as follows:

\textbf{Low-Frequency Components}. The low-frequency components of the target image $I_{SR}$ can be formulated as a function of the low-resolution image $I_{LR}$ and the upscaling process. Mathematically, the ideal low-frequency components $A_{SR}$ of the SR image can be expressed as:

\begin{equation}
\label{fuc:22}
A_{SR} \sim p(A_{SR} | H(I_{LR})) + \epsilon
\end{equation}

where $H$ is the probability density transformation function which maps $I_{LR}$ to the domain of $A_{SR}$, and independent Gaussian noise $\epsilon$ accounts for uncertainties and imperfections in the transformation process. The output of $H(I_{LR})$ denotes as a condition that constraints the probability flow of $A_{SR}$ during the UR process. The low frequency components obeys the conditional probability distribution $p(A_{SR} | H(I_{LR}))$.   \par
However, due to the ill-posed nature of the transformation function, the analytic form of $H$ in Eq. \ref{fuc:22}   is hard to achieve. Consequently, we use a network $\tau_{\theta}$ to estimate the transformation function. 

\begin{equation}
	\label{fuc:281}
	\tau_{\theta}(I_{LR}) = SR(p(I_{LR})) + \epsilon \rightarrow \mathbb{R}^{A_{SR}}
\end{equation}

where $SR(p(I_{LR}))$ can be any arbitrary pre-trained SR pipeline to provide the map of $I_{LR}$: $\mathbb{R}^{I_{LR}} \rightarrow \mathbb{R}^{A_{SR}}$. Mathematically,  $\tau_{\theta}(I_{LR})$ is a projection-based condition for alleviating the ill-posed problem. Thus, $A_{SR}$ can be generated based on the conditional probability distribution, $p(A_{SR}|\tau_{\theta})$. The low-frequency SR process is further formulated as a reverse diffusion SDE:

\begin{equation}
	\label{fuc:210}
	dA_{SR} = \left[\bar{f}(A_{SR, t}, t) - \bar{g}(t)^2 \nabla_{A_{SR, t}} \log p_t(A_{SR, t} \mid \tau_{\theta}(I_{LR}))\right] dt + \bar{g}(t) d\bar{w}
\end{equation}

The conditional score function, $\nabla_{I_{SR}} \log p(I_{SR}|\tau_{\theta}(I_{LR}))$, requires model retraining whenever the cross-scale condition $\tau_{\theta}(I_{LR})$ is updated. This dependency limits the generalization capability of the low-frequency SR process. To address this issue, inspired by \cite{chung2022improving}, we adopt an unconditional score function, $\nabla_x \log p_t(A_{SR})$, independent of $\tau_{\theta}(I_{LR})$. Instead, $\tau_{\theta}(I_{LR})$ is incorporated as an additional input or condition in the score-based framework, which guides the denoising process without constraining the model's adaptability. More specifically, it can be described as follows:

\begin{equation}
	\label{fuc:23}
	A'_{SR, t-1} = f(A_{SR, t}, s_{\theta}) + g(A_{SR, t}) \cdot z, \quad z \sim N(0, I)
\end{equation}

\begin{equation}
	\label{fuc:24}
	A_{SR, t-1} = \alpha \cdot A'_{SR, t-1} + b_t
\end{equation}

where $\alpha$, $b_i$ are functions of $\tau_{\theta}(I_{LR})$ and $A_{SR,t}$. Note that Eq. \ref{fuc:23} is identical to the unconditional reverse diffusion item in Eq. \ref{fuc:3}, whereas Eq. \ref{fuc:24} includes the projection-based condition. \par

\textbf{High-frequency Wavelet Components}. In WaveDiffUR, the up-scaler model is tasked to transform the low-resolution wavelet component  $VHD_{LR} = \{V_{LR}, H_{LR}, D_{LR}\}$ into their high-resolution counterparts $VHD_{SR} = \{V_{SR}, H_{SR}, D_{SR}\}$. The process is formulated as an up-scaling transform followed by the addition of independent Gaussian noise. The up-scaler generates $VHD_{SR}$ as:

\begin{equation}
\label{fuc:2112}
VHD_{SR} \sim \mathcal{N} \left( \mathbb{U}(VHD_{LR}), r^2 \mathbb{U} \mathbb{U}^\top + \sigma^2 I \right)
\end{equation}

where $\mathbb{U}$ is an upscale function, $r$ is a standard deviation value associated with $VHD_{LR}$, and $\sigma^2 I$ is independent Gaussian noise.The upscale function $\mathbb{U}$ can be estimated using a pre-trained SR model. Thus, Equation \ref{fuc:2112} is rewritten as:

\begin{equation}
	\label{fuc:2113}
	VHD_{SR} \sim \mathcal{N} \left( \mathbb{U}_{SR}(VHD_{LR}), r^2 \mathbb{U}_{SR} \mathbb{U}_{SR}^\top + \sigma^2 I \right)
\end{equation}

The detailed information on the WaveDiffUR method is shown in Algorithm \ref{algo:0}.

\begin{algorithm}
\caption{WaveDiffUR SDE solver}
\label{algo:0}
\begin{algorithmic}[1]
\State \textbf{input}: low-resolution image $I_{LR}$, pre-trained SR pipeline $SR$, pre-trained Unet $s_{\theta}$, current resolution $r$, UR resolution $R$, SR rate $k$.
\State UR rate $K = R/r$, UR step $d = K/k$
\State Perform a DWT
\State $\{A_{LR}, VHD_{LR}\} = DWT(I_{LR})$
\State $\tau_{\theta}(I_{LR}) = SR(p(I_{LR})) + \epsilon$
\For{$i$ = 1 to $d$}
    \For{$t$ = $T$ to 1}
        \State low-frequency wavelet SR
        \State $A'_{SR, t-1} = f(A_{SR, t}, s_{\theta}) + g(A_{SR, t}) \cdot z$
        \State $A_{SR, t-1} = \alpha \cdot A'_{SR, t-1} + b_i$    
    \EndFor
    \State high-frequency wavelet restoration
    \State $VHD_{SR} \sim \mathcal{N} \left( \mathbb{U}_{SR}(VHD_{LR}), r^2 \mathbb{U}_{SR} \mathbb{U}_{SR}^\top + \sigma^2 I \right)$
    \State $I_{SR} = IDWT(\{A_{SR, 0}, VHD_{SR}\})$
    \While{$i \neq d$}
        \State $I_{LR} = I_{SR}$
    \EndWhile
    \While{$i = d$}
        \State $I_{UR} = I_{SR}$ 
    \EndWhile
\EndFor
\State \textbf{output}: $I_{UR}$
\end{algorithmic}
\end{algorithm}

\subsection{Cross-scale pyramid (CSP) constraint WaveDiffUR solver}

Considering the degradation issue of re-utilizing fixed boundary conditions $\tau_{\theta}(I_{LR})$ in constraining the WaveDiffUR SDE, we propose a dynamically updated boundary condition, named cross-scale pyramid (CSP), to better constraint the ill-posed problem and improve the cross-scale inference capability of $\tau_{\theta}$. To this end, we introduce a reference image $I_{ref} \in \mathbb{R}^{H' \times W' \times c}$, $H' \geq H$, $W' \geq W$ into our study, and the CSP condition $\tau_{\theta}(I_{LR}, I_{ref})$ can be modeled with joint probability $p(I_{ref}$, $I_{LR})$: 

\begin{equation}
\label{fuc:28}
\tau_{\theta}(I_{LR}, I_{ref}) = T_{\theta} \cdot p(I_{ref}, I_{LR}) + \epsilon \rightarrow \mathbb{R}^{A_{SR}}
\end{equation}

$\tau_{\theta}(I_{LR}, I_{ref})$, abbreviated as $\tau^{lf}_{\theta}$, ensures that the domain of the conditions at each UR level remains close to the domain of the step-wised SR target. This alignment facilitates the ill-posed inverse process in accordance with the cross-scale unmixing rule. Therefore, $\tau^{lf}_{\theta}$ is employed as a conditional constraint to solve the inverse diffusion problem at each pyramid level. The low-frequency wavelet components of SR are expressed as:

\begin{equation}
\label{fuc:29}
A_{SR,t} \sim p(A_{SR,t} | \tau^{lf}_{\theta}) + \epsilon
\end{equation}

It is important to note that the parameter $\theta$ evolves progressively with the increasing magnification rate, thereby enhancing the efficiency of the conditional constraint for the ill-posed inverse diffusion UR problem. Once the form of $\tau^{lf}_{\theta}$ is identified, the fixed condition $\tau_{\theta}(I_{LR})$ should be replaced with it. Accordingly, the diffusion SDE for low-frequency wavelet component of SR can be solved using Eqs. \ref{fuc:210} - \ref{fuc:24}. \par

To restore the sparse representations of high-frequency wavelet components ($VHD_{SR}$) in SR images with fidelity comparable to the HR images($VHD_{HR}$), the CSP-WaveDiffUR solver employs Cross-scale Pyramid (CSP) constraints. These constraints ensure that the upscaling and inter-band interactions are guided by both the low-resolution input ($VHD_{LR}$) and the reference information ($VHD_{ref}$). The high-frequency CSP constraints are formulated in terms of $VHD_{LR} = \{V_{LR}, H_{LR}, D_{LR}\}$ and  $VHD_{ref} = \{V_{ref}, H_{ref}, D_{ref}\}$ as follows:

\begin{equation}
\label{fuc:321}
\tau^V_{\theta} = T_{\theta_V} \cdot p(V_{LR}, V_{ref})
\end{equation}

\begin{equation}
\label{fuc:322}
\tau^H_{\theta} = T_{\theta_H} \cdot p(H_{LR}, H_{ref})
\end{equation}

\begin{equation}
\label{fuc:323}
\tau^D_{\theta} = T_{\theta_D} \cdot p(D_{LR}, D_{ref})
\end{equation}

where $\tau^{(.)}_{\theta}$ are constraints for vertical (V), horizontal (H)  and diagonal (D) details; $T_{\theta{(.)}}$ are transformations for the respective sub-bands and $p(.)$ are cross-scale dependency modeled between low-resolution input and reference wavelet components.  These constraints enhance spectral-spatial consistency by aligning the upscaled components with both the low-resolution input and reference information.    
The high-frequency wavelet component $VHD_{SR}$ are restored using the CSP constraints: 

\begin{equation}
\label{fuc:32}
VHD_{SR} \sim \mathcal{N} \left( \mathbb{M}(VHD_{LR}, \tau^{hf}_{\theta}), r^2 \mathbb{M} \mathbb{M}^\top + \sigma^2 I \right)
\end{equation}

where $\mathbb{M}$ is a high-frequency restoration function, which predicts sparse representations of vertical(V), horizontal(H), and diagonal (D) components of the SR image under the CSP constraints $\tau^{hf}_{\theta} = \{\tau^V_{\theta}, \tau^H_{\theta}, \tau^D_{\theta}\}$, $r^2$ is current resolution in high-frequency restoration. The details of $\mathbb{M}$ are described in Section \ref{sec:3.34}. The full details about the CSP-WaveDiffUR solver is outlined in Algorithm \ref{algo:1}. The overview of CSP-WaveDiffUR is shown in Fig. \ref{fig:1w}. \par

\begin{algorithm}
\caption{CSP-WaveDiffUR SDE solver}
\label{algo:1}
\begin{algorithmic}[1]
\State \textbf{input}: low-resolution image $I_{LR}$, reference image $I_{ref}$, pre-trained Unet $s_{\theta}$, CSP encoder $T_{\theta}$, high-frequency restoration module $\mathbb{M}$, current resolution $r$, UR resolution $R$, SR rate $k$.
\State UR rate $K = R/r$, UR step $d = K/k$
\State Perform a DWT
\State $\{A_{LR}, VHD_{LR}\} = DWT(I_{LR})$
\State $\tau^{lf}_{\theta}\ = T_{\theta} \cdot p(I_{ref}, I_{LR}) + \epsilon$
\State $\tau^{hf}_{\theta}\ = T_{\theta} \cdot p(hf_{ref}, hf_{LR}) + \epsilon$
\For{$i$ = 1 to $d$}
    \For{$t$ = $T$ to 1}
        \State low-frequency wavelet SR
        \State $A'_{SR, t-1} = f(A_{SR, t}, s_{\theta}) + g(A_{SR, t}) \cdot z$
        \State $A_{SR, t-1} = \alpha \cdot A'_{SR, t-1} + b_i$    
    \EndFor
    \State high-frequency wavelet restoration
    \State $VHD_{SR} \sim \mathcal{N} \left( \mathbb{M}(VHD_{LR}, \tau^{hf}_{\theta}), r^2 \mathbb{M} \mathbb{M}^\top + \sigma^2 I \right)$
    \State $I_{SR} = IDWT(\{A_{SR, 0}, VHD_{SR}\})$
    \While{$i \neq d$}
        \State $I_{LR} = I_{SR}$
    \EndWhile
    \While{$i = d$}
        \State $I_{UR} = I_{SR}$ 
    \EndWhile
\EndFor
\State \textbf{output}: $I_{UR}$
\end{algorithmic}
\end{algorithm}

\begin{figure}[]   
    \centering  
    \includegraphics[width=5.5in]{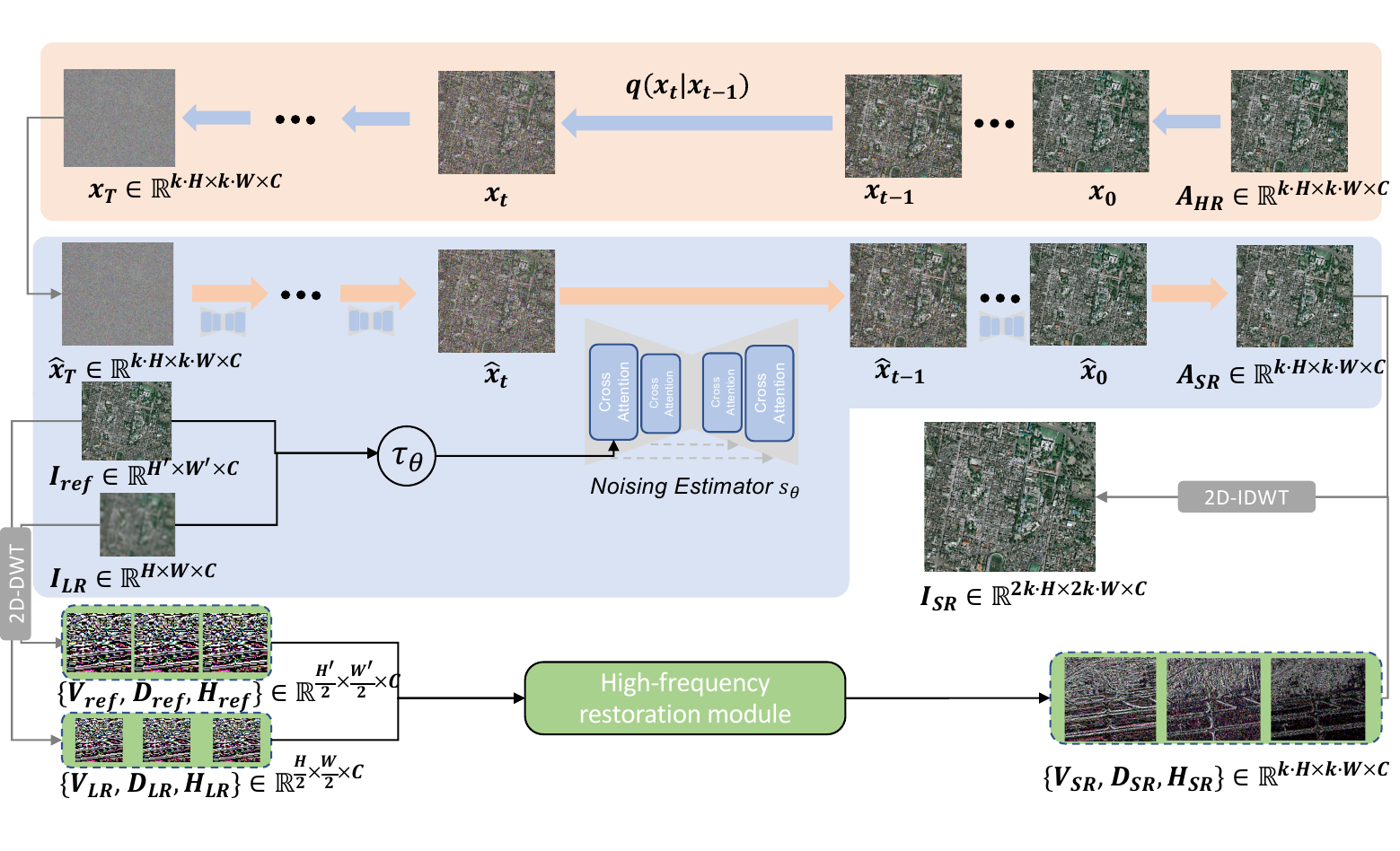}  
    \caption{An overall of the proposed CSP-WaveDiffUR.}
    \label{fig:1wa}  
\end{figure}

\subsubsection{Cross-scale pyramid encoder}
\label{sec:3.3}
\begin{figure}[]   
    \centering  
    \includegraphics[width=4in]{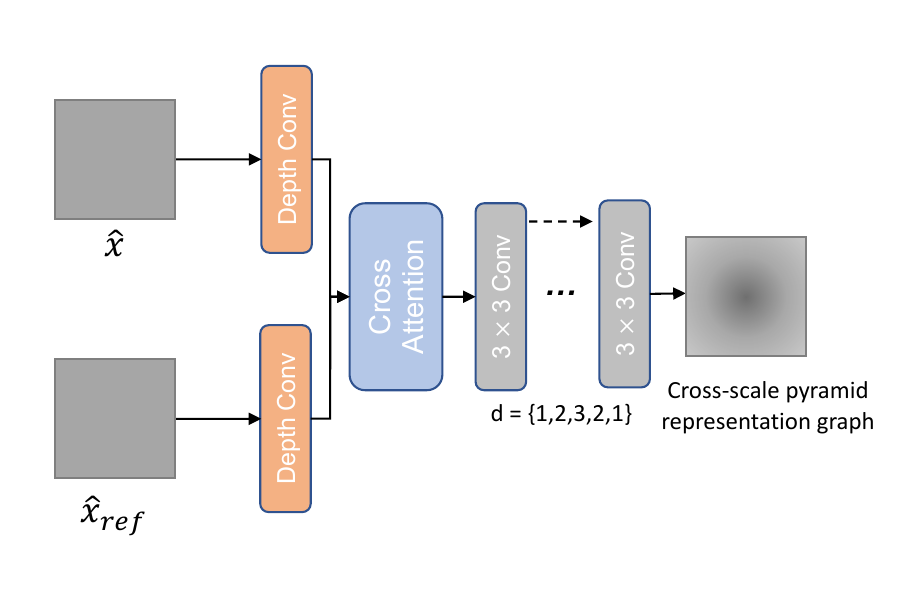}  
    \caption{An illustration of the cross-scale pyramid encoder. }
    \label{fig:4w}  
\end{figure}

The Cross-Scale Pyramid Encoder (CSP Encoder) is a critical module in the CSP-WaveDiffUR framework, designed to dynamically generate cross-scale constraints for spectral-spatial unmixing. Fig. \ref{fig:4w} illustrates the structure of the CSP encoder that models the spectral-spatial unmixing rule between a given input ($(\hat{x}$) and its corresponding reference ($\hat{x}_{ref}$).

\begin{comment}

The primary motivation of this work is to construct a CSP constraint from the given input $\hat{x}$ and the corresponding reference $\hat{x}_{ref}$, which pave the way to solve the ill-pose diffusion SR process. To this end, we develop a cross-attention-driven auto-encoder $\tau_{\theta}(\hat{x}, \hat{x}_{ref})$, as shown in Fig. \ref{fig:4w} that models the spectral-spatial unmixing rule between the given input pair $(\hat{x}, \hat{x}_{ref})$. 
\end{comment}

To efficiently extract the features of inputs, we use depth-wise separable convolutions \cite{chollet2017xception} in the encoder. After that, the extracted features are fed into cross-attention layers to compute an intermediate representation tensor. The cross-attention mechanism leverages the information in both $\hat{x}$ and $\hat{x}_{ref}$, effectively modeling the joint probability distribution $p(\hat{x} \mid \hat{x}_{ref})$. The intermediate representation tensor is computed as:

\begin{equation}
\label{fuc:3.6}
Attention(Q, K, V) = softmax (\frac{Q \cdot K^{\tau}}{\sqrt{d}} \cdot V)
\end{equation}

where, $Q = W^{(i)}_Q \cdot \phi_i(\hat{x}_{SR, t})$, $K = W^{(i)}_K \cdot \tau_\theta(\hat{x}, \hat{x}_{ref})$, $V = W^{(i)}_V \cdot \tau_\theta(\hat{x}, \hat{x}_{ref})$, in which $\phi_i(\hat{x}_{SR, t})$ is a flattened latent space features generated from a DeepConv block, $W^{(i)}_Q$, $W^{(i)}_K$, and $W^{(i)}_V$ are learnable projection matrices. \par
To enhance the representation of cross-scale spectral-spatial information, a progressive dilation Resblock is incorporated into the encoder. In this block, the first and last convolutions are utilized for local feature extraction, and a sequence of dilation convolutions are applied to expand the receptive field, capturing cross-scale dependencies. \par

\subsubsection{High-frequency restoration module}
\label{sec:3.34}
\begin{figure}[]   
    \centering  
    \includegraphics[width=5in]{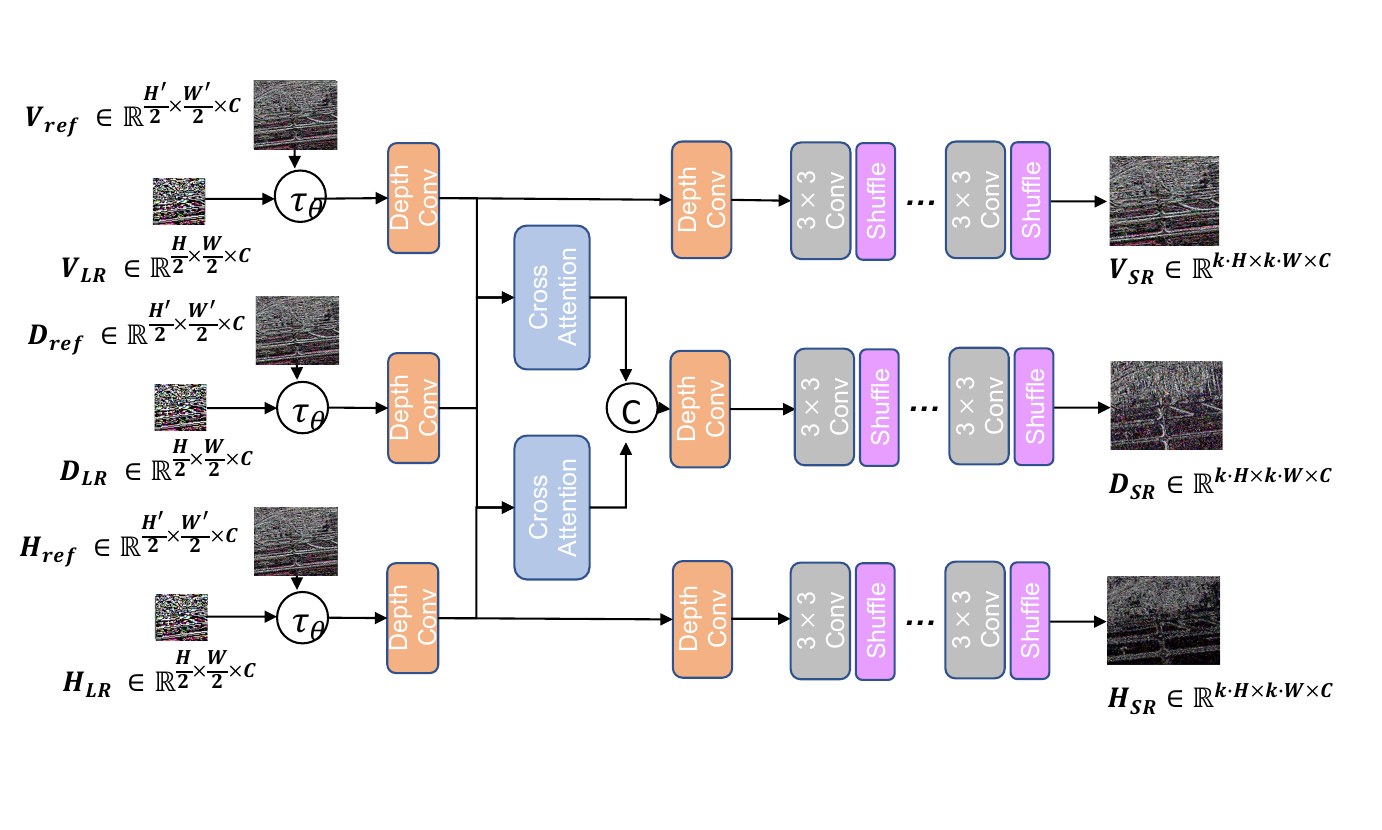}  
    \caption{An illustration of the cross-scale high-frequency restoration model. }
    \label{fig:4h}  
\end{figure}

The cross-scale high-frequency restoration (CSHR) module  ( Fig. \ref{fig:4h} ) in CSP-WaveDiffUR is designed to reconstruct the high-frequency wavelet coefficients $\{V_{SR}, H_{SR}, D_{SR}\}$ from their low-resolution counterparts $\{V_{LR}, H_{LR}, D_{LR}\}$ and reference coefficients $\{V_{ref}, H_{ref}, D_{ref}\}$. This reconstruction enhances the vertical, horizontal, and diagonal details of the target SR image by effectively modeling cross-scale spectral-spatial interactions. The architecture of the cross-scale high-frequency restoration (CSHR) module is shown in Fig. \ref{fig:4h}. The module comprises the following steps: (1) Feature extraction with depth-wise separable convolutions from the input coefficients efficiently, (2) Modeling the interaction of information in $V$ and $H$ that augment the details in $D$ using two cross-attention layers \cite{hou2019cross}, (3) Generating SR high-frequency coefficients with a progressive dilation ResBlock developed under inspiration from the work of Hai \textit{et al} \cite{hai2022combining}, which includes a dilation convolution followed by a shuffle layer for up-scaling of high-frequency coefficients.

\subsection{Model training}
In addition to the objective function $L_{diff}$ utilized for optimizing the diffusion model, a high-frequency realness loss $L_{realness}$ is employed, which combines Mean Squared Error (MSE) loss and Total Variation (TV) loss, similar to the approach described by Liu \textit{et al.} \cite{liu2021retinex}. This loss function is employed to reconstruct the high-frequency coefficients.

\begin{equation}
\label{fuc:3.7}
L_{realness} = \lambda_1 \| \{V_{SR}, H_{SR}, D_{SR}\} - \{V_{H}, H_{H}, D_{H}\} \|^2 + \lambda_2 TV(\{V_{SR}, H_{SR}, D_{SR}\})
\end{equation}

where $\lambda_1 = 0.1$ and $\lambda_2 = 2$ are weights for the respective terms term. Moreover, we incorporate a consistent loss $L_{consistent}$ that combines $L1$ loss and SSIM loss to ensure the fidelity of the reconstructed super-resolution image $I_{SR}$ compared to the ground truth high-resolution image $I_{H}$.

\begin{equation}
\label{fuc:3.8}
L_{consistent} = \| I_{SR} - I_{H} \| + (1 - \text{SSIM}(I_{SR} - I_{H})).
\end{equation}

The total loss for the proposed neural network can be expressed as:

\begin{equation}
\label{fuc:3.9}
L_{total} = L_{diff} + L_{realness} +L_{consistent} 
\end{equation}

\section{Experiment and discussion}
\label{sec:ed}

\subsection{Dataset}
To comprehensively evaluate the effectiveness of the methods presented in this study, we trained and fine-tuned our model on two publicly available datasets (ImageNet 1K \cite{russakovsky2015imagenet}, and AID \cite{xia2017aid}). In addition, we evaluated the model performance on two public datasets (DOTA \cite{xia2018dota} and DIOR \cite{li2020object}) and one self-designed dataset (Dataset for drone images super-resolution (DDISR)) \cite{shi2022latent}). \par

The details of the datasets for model training, testing and evaluation are detailed in Table \ref{tab:0_1}. In the pre-training phase, ImageNet 1K was used for ($64 \times 64 \rightarrow 128 \times 128$) super resolution tasks and the dev sets were used for validation. Wherein, the original images were respectively resized to $64 \times 64$ (low-resolution), $96 \times 96$ (reference), $128 \times 128$ (high-resolution). In the fine-tuning phase,  AID dataset were used for ($64 \times 64 \rightarrow 256 \times 256$) remote sensing super resolution tasks. The AID training set consisted of 8000 randomly selected images, with the remaining 2000 images as the testing set. Each image had the counterpart resolution of $64 \times 64$, $96 \times 96$, $256 \times 256$. In the model testing phase, the subsets of images from the public accessible DOTA dataset and the DIOR dataset were used for model evaluation. It contains a total of 1080 images, 480 from DOTA and 600 from DIOR (30 randomly selected per class). In the data pre-processing phase, we employed bicubic interpolation for image degradation.\par

In real-world analysis with large scale factors, we employed the WWDSSO dataset. It consists of 300 drone-satellite synchronous observation pairs of Landset-8 image with 30m resolution ( as low-resolution input), Sentinel-2 image with 10m resolution (as mid-resolution reference), and Drone image with 0.23m resolution (as ground truth). The super-resolution performance was evaluated for three scale factors $\times 10$, $\times 20$, $\times 100$. The simulated degradation was applied on ground truth drone images for scale factors ranging from $\times 2$ to $\times 128$ super-resolution.  

\begin{table}[]

\caption{The details of the datasets for model training, testing, and evaluation.}
\label{tab:0_1}

\centering
\resizebox{4in}{!}{
\begin{tabular}{cccccc}
\toprule
Data type  & ImageNet  & AID   & DOTA & DIOR & DDISR       \\\bottomrule
Scale for  & $\times 2$& $\times 2$, $\times 4$   & $\times 2$, $\times 4$  & $\times 2$, $\times 4$  & $\times 2$ to $\times 128$ \\
Classes    & 1000      & 30    & 16   & 20   & 1           \\
Pre-train  & 1,281,167 &       &      &      &             \\
Fine-tune  &           & 8,000 &      &      &             \\
Testing    & 500,000   & 2,000 &      &      &             \\
Evaluation &           &       & 480  & 600  & 300    \\    \bottomrule      
\end{tabular}
}
\end{table}

\subsection{Metrics}
In this study, seven metrics were employed to comprehensively evaluate the performance of the super-resolution model. These included three full-reference metrics for assessing the similarity between super-resolution and ground-truth images: Fréchet Inception Distance (FID) \cite{heusel2017gans}, the widely used Peak Signal-to-Noise Ratio (PSNR), and Structural Similarity Index (SSIM) \cite{wang2004image}. Among these, FID was extensively used to measure the generative quality of the model, improving upon the Inception Score (IS) \cite{xiao2023ediffsr} by directly measuring feature-level distances without relying on a classifier.
To assess pixel-wise spectral fidelity, two metrics—Spectral Angle Mapper (SAM) and Spectral Reconstruction Error (SRE) \cite{shi2022latent}—were utilized to calculate the average angle between super-resolution images and ground-truth data. Additionally, two reference-free metrics were included: the Natural Image Quality Evaluator (NIQE) \cite{mittal2012making}, which measures perceptual quality, and Average Gradient (AG), which evaluates high-frequency detail preservation. These metrics provided a holistic evaluation of the model’s performance, addressing quantitative accuracy (PSNR, SSIM), perceptual quality (FID, NIQE), spectral consistency (SAM, SRE), and sharpness (AG).

\subsection{Implementation Details}
The proposed model was implemented with PyTorch. The model training was conducted on three NVIDIA RTX 2080Ti GPUs. The network underwent pre-training for $1 \times 10^4$ iterations on the ImageNet 1K training dataset, and fine-tinning was performed for $1 \times 10^3$ iterations on the AID training dataset. The Adam optimizer \cite{kingma2014adam} was utilized for optimization, with an initial learning rate of $1 \times 10^{-4}$, which decayed by a factor of 0.8 every $5 \times 10^3$ iterations. The batch size set to 8, and the patch size was 256×256. The wavelet transformation scale $K$ was set to $4$ during pre-training and $8$ during fine-tuning. The U-Net architecture \cite{ronneberger2015u} was employed as the noise estimator network in the diffusion process. To achieve efficient super-resolution, the time step $T$ was set to 1000 during the training phase.

\subsection{Model evaluation on SR tasks}
\label{sec:43}
To assess the performance of the proposed model, we conducted a  comparative analysis. The comparative analysis included our model and several  state-of-the-art (SOTA) super-resolution approaches including LE-GAN \cite{shi2022latent}, ESRGAN-DP \cite{song2023esrgan}, RefVAE \cite{liu2021variational}, LSR-SR \cite{guo2022lar}, SR3 \cite{saharia2022image}, IRSDE \cite{luo2023image} EDiffSR \cite{xiao2023ediffsr}, LWTDM \cite{an2023efficient}. These SOTA SR models are pre-dominant techniques within the fields, representing diverse methodologies. Specifically, LE-GAN and ESRGAN-DP represent GAN-based methods. RefVAE and LSR-SR represent VAE-based image super-resolution methods. Conversely, SR3 and IRSDE are cutting-edge diffusion-based models for natural image super-resolution. EDiffSR and LWTDM are cutting-edge diffusion-based models specific to remote sensing image super-resolution. All these comparative models were fine-tuned on the AID training dataset according to the configurations specified in their official implementations, ensuring a fair and consistent comparison.

\textit{Quantitative Comparison.} In the study presented, the average values of FID, PSNR, SSIM, SAM, NIQE and AG across the ImageNet and AID test sets are list in Table \ref{tab:2}. It is observed that, for the ImageNet testing dataset, the proposed model achieves top-tier performance, ranking either $1^{st}$  or showing negligible difference ( $<0.1\%$) from the top SOTA moddels (e.g. SR3 and EDiffSR) in FID, PSNR, and SSIM. It is noteworthy that GAN-based models, such as LE-GAN, generally excel in achieving high channel realness scores, largely due to the alignment between spectral loss computation in adversarial feature space and channel realness assessment methods, particularly for SRE. The proposed method outperforms diffusion-based models, achieving $13.12\%$ higher SAM score than SR3. In terms of reference-free metrics (i.e. NIQE and AG), our proposed model demonstrates the best performance, underscoring its ability to preserve structural integrity in low-frequency components.\par

The similar results can be found when testing on the AID test dataset with a $\times 2$ scale factor. The performance of the proposed model matches EDiffSR in FID, PSNR and SSIM, is comparable with LE-GAN in SAM and SRE, and superior in NIQE and AG. For the $\times 2$ scale factor, the proposed model achieves the minimal degradation in restoration scores. The highest overall performance across all metrics demonstrates its exceptional ability to restore accurate structural details in remote sensing image super-resolution. Best NIQE and AG scores indicate the results closely align with human visual perception. \par 

\begin{table}[]

\caption{A quantitative comparison of the SOTA SR models pre-trained on the ImageNet dataset ($\times 2$) and fine-tuned on the AID dataset ($\times 2$, and $\times 4$) in terms of FID, PSNR, SSIM, SAM, SRE, NIQE, and AG. The tests are conducted on the dev split test data. The best values are highlighted }
\label{tab:2}
\centering
\resizebox{6in}{!}{
\begin{tabular}{cccccccccc}
\toprule
\multicolumn{10}{c}{ImageNet}                                                           \\\midrule
\multicolumn{1}{l}{Category} & Method    & \multicolumn{1}{l}{Scale} & FID $\downarrow$   & PSNR$\uparrow$  & SSIM$\uparrow$ & SAM $\downarrow$  & SRE$\downarrow$   & NIQE $\downarrow$  & AG $\uparrow$    \\
\multirow{2}{*}{GAN}         & LE-GAN    & \multirow{9}{*}{$\times 2$}  & 48.39 & 48.27 & 0.85 & \textbf{5.61}  & \textbf{5.16}  & 12.82 & 4.51 \\
                             & ESRGAN-DP &                           & 53.26 & 48.02 & 0.84 & 6.97  & 8.35  & 12.24 & 4.81 \\
\multirow{2}{*}{VAE}         & RefVAE    &                           & 47.15 & 45.61 & 0.8  & 7.66  & 9.04  & 13.27 & 4.98 \\
                             & LSR-SR    &                           & 47.44 & 45.49 & 0.84 & 7.84  & 8.28  & 12.84 & 5.12 \\
\multirow{5}{*}{Diffusion}   & SR3       &                           & 40.52 & 50.47 & \textbf{0.95} & 6.55  & 7.32  & 10.48 & 4.82 \\
                             & IRSDE     &                           & 45.03 & 48.56 & 0.85 & 6.62  & 7.94  & 12.42 & 5.36 \\
                             & EDiffSR   &                           & \textbf{40.41} & 49.85 & 0.88 & 6.68  & 7.17  & 11.04 & 5.62 \\
                             & LWTDM     &                           & 43.98 & 46.3  & 0.77 & 7.8   & 9.07  & 14.27 & 4.33 \\
                             & Proposed  &                           & 40.45 & \textbf{52.97} & \textbf{0.95} & 5.79  & 5.65  & \textbf{9.86}  & \textbf{5.85} \\
\midrule
\multicolumn{10}{c}{AID}                                                                                              \\
\midrule
\multicolumn{1}{l}{Category} & Method    & \multicolumn{1}{l}{Scale} & FID $\downarrow$   & PSNR$\uparrow$  & SSIM$\uparrow$ & SAM $\downarrow$  & SRE$\downarrow$   & NIQE $\downarrow$  & AG $\uparrow$   \\
\multirow{2}{*}{GAN}         & LE-GAN    & \multirow{9}{*}{$\times 2$}        & 53.13 & 39.8  & 0.71 & 8.09  & \textbf{7.68}  & 14.3  & 3.9  \\
                             & ESRGAN-DP &                           & 55.2  & 39.62 & 0.7  & 9.66  & 10.71 & 16.08 & 3.91 \\
\multirow{2}{*}{VAE}         & RefVAE    &                           & 59.76 & 37.7  & 0.67 & 10.39 & 11.46 & 17.01 & 3.73 \\
                             & LSR-SR    &                           & 60.21 & 37.22 & 0.7  & 10.58 & 11.37 & 17.06 & 3.65 \\
\multirow{5}{*}{Diffusion}   & SR3       &                           & 51.43 & 40.8  & 0.67 & 10.07 & 10.31 & 13.2  & 3.98 \\
                             & IRSDE     &                           & 53.8  & 40.41 & 0.71 & 9.35  & 10.46 & 15.7  & 4.06 \\
                             & EDiffSR   &                           & 51.49 & \textbf{41.49} & 0.73 & 9.11  & 9.99  & 14.98 & 4.07 \\
                             & LWTDM     &                           & 55.06 & 39.56 & 0.64 & 10.72 & 12.03 & 18.04 & 3.65 \\
                             & Proposed  &                           & \textbf{50.9}  & 41.48 & \textbf{0.75} & \textbf{8.07}  & 7.71  & \textbf{12.03} & \textbf{4.24} \\
\midrule
\multirow{2}{*}{GAN}         & LE-GAN    & \multirow{9}{*}{$\times 4$}  & 56.19 & 35.14 & 0.7  & 10.14 & 9.12  & 16.32 & 3.46 \\
                             & ESRGAN-DP &                           & 61.05 & 35.64 & 0.63 & 11.31 & 12.75 & 18.67 & 2.82 \\
\multirow{2}{*}{VAE}         & RefVAE    &                           & 66.04 & 33.82 & 0.6  & 12.01 & 13.24 & 19.26 & 3.35 \\
                             & LSR-SR    &                           & 66.25 & 32.97 & 0.63 & 11.79 & 13.12 & 19.04 & 3.17 \\
\multirow{5}{*}{Diffusion}   & SR3       &                           & 66.02 & 35.85 & 0.6  & 11.51 & 11.84 & 14.86 & 3.4  \\
                             & IRSDE     &                           & 59.58 & 35.75 & 0.64 & 11.22 & 11.82 & 17.36 & 2.98 \\
                             & EDiffSR   &                           & 56.67 & 37.17 & 0.66 & 10.06 & 11.05 & 16.8  & 3.55 \\
                             & LWTDM     &                           & 67.34 & 32.23 & 0.58 & 12.63 & 14.11 & 20.11 & 2.52 \\
                             & Proposed  &                           & \textbf{53.18} & \textbf{38.64} & \textbf{0.71} & \textbf{8.97}  & \textbf{8.47}  & \textbf{13.75} & \textbf{4.14}\\
                             \bottomrule
\end{tabular}
}
\end{table}

Additional results for assessing the generalization capacity of the proposed model on independent datasets are shown in Table \ref{tab:3}, where the highest performances are highlighted in bold. The evaluation results indicate that the proposed model achieves the first place across 6 out of 7 metrics on the $\times 2$ DOTA dataset. It ranks first in all metrics on $\times 4$ DOTA dataset, and $\times 2$ and $\times 4$ DIOR datasets. The results for both $\times 2$ and $\times 4$ scale factors display  consistent performance. The minimal performance variation across scale factors highlights the model’s robustness in handling varying resolutions. These findings confirm the proposed model’s excellent generalization capacity.

\begin{table}[]

\caption{A quantitative comparison of the generalized capability between the SOTA SR models on independent DOTA and DIOR datasets for $\times 2$ and and $\times 4$ super-resolution tasks in terms of FID, PSNR, SSIM, SAM, SRE, NIQE, and AG. The best values are highlighted in bold. }
\label{tab:3}

\centering
\resizebox{6in}{!}{
\begin{tabular}{cccccccccc}
\toprule
\multicolumn{10}{c}{DOTA}                                                                                                  \\\midrule
\multicolumn{1}{l}{Category} & Method    & \multicolumn{1}{l}{Scale} & FID $\downarrow$   & PSNR$\uparrow$  & SSIM$\uparrow$ & SAM $\downarrow$  & SRE$\downarrow$   & NIQE $\downarrow$  & AG $\uparrow$    \\
\multirow{2}{*}{GAN}         & LE-GAN    & \multirow{9}{*}{$\times 2$} & 26.34 & 21.31 & 0.78 & \textbf{6.48}  & 7.04  & 14.68 & 2.14 \\
                             & ESRGAN-DP &                           & 26.89 & 21.69 & 0.73 & 9.81  & 10.78 & 17.43 & 3.48 \\
\multirow{2}{*}{VAE}         & RefVAE    &                           & 29.65 & 20.27 & 0.81 & 10.99 & 13.27 & 18.67 & 2.28 \\
                             & LSR-SR    &                           & 29.46 & 20.26 & 0.8  & 12.51 & 12.73 & 18.22 & 2.09 \\
\multirow{5}{*}{Diffusion}   & SR3       &                           & 24.93 & 21.82 & 0.79 & 11.51 & 10.81 & 14.42 & 2.73 \\
                             & IRSDE     &                           & 26.82 & 21.9  & 0.7  & 11.17 & 11.96 & 16.63 & 3.07 \\
                             & EDiffSR   &                           & 24.92 & 22.66 & 0.71 & 9.21  & 11.03 & 16.77 & 2.58 \\
                             & LWTDM     &                           & 27.48 & 21.16 & 0.74 & 11.65 & 13.22 & 19.97 & 1.9  \\
                             & Proposed  &                           & \textbf{20.63} & \textbf{28.53} & \textbf{0.93} & 6.58  & \textbf{6.61}  & \textbf{10.18} & \textbf{6.23} \\
\multirow{2}{*}{GAN}         & LE-GAN    & \multirow{9}{*}{$\times 4$} & 26.72 & 19.55 & 0.72 & 11.86 & 11.41 & 19.37 & 2.06 \\
                             & ESRGAN-DP &                           & 31.29 & 20.54 & 0.67 & 14.06 & 14.4  & 20.78 & 3.45 \\
\multirow{2}{*}{VAE}         & RefVAE    &                           & 34.52 & 20.15 & 0.72 & 14.57 & 16.3  & 23.04 & 2.24 \\
                             & LSR-SR    &                           & 32.98 & 17.94 & 0.71 & 15.27 & 15.62 & 21.65 & 2.04 \\
\multirow{5}{*}{Diffusion}   & SR3       &                           & 27.44 & 20.43 & 0.72 & 13.07 & 15.46 & 18.05 & 2.69 \\
                             & IRSDE     &                           & 27.48 & 19.62 & 0.62 & 13.15 & 16.21 & 20.96 & 3.06 \\
                             & EDiffSR   &                           & 27.75 & 22.34 & 0.68 & 14.22 & 13.22 & 20.84 & 2.55 \\
                             & LWTDM     &                           & 30.31 & 19.87 & 0.59 & 15.2  & 14.15 & 22.34 & 1.87 \\
                             & Proposed  &                           & \textbf{22.65} & \textbf{28.21} & \textbf{0.91} & \textbf{7.27}  & \textbf{7.82}  & \textbf{11.98} & \textbf{6.17} \\\midrule
\multicolumn{10}{c}{DIOR}                                                                                                  \\\midrule
\multicolumn{1}{l}{Category} & Method    & \multicolumn{1}{l}{Scale} & FID $\downarrow$   & PSNR$\uparrow$  & SSIM$\uparrow$ & SAM $\downarrow$  & SRE$\downarrow$   & NIQE $\downarrow$  & AG $\uparrow$  \\
\multirow{2}{*}{GAN}         & LE-GAN    & \multirow{9}{*}{$\times 2$} & 26.09 & 22.71 & 0.79 & 10.11 & 11.75 & 18.97 & 5.41 \\
                             & ESRGAN-DP &                           & 30.56 & 18.46 & 0.65 & 11.17 & 18.06 & 18.66 & 6.37 \\
\multirow{2}{*}{VAE}         & RefVAE    &                           & 30.59 & 17.04 & 0.62 & 12.67 & 24.37 & 20.51 & 2.62 \\
                             & LSR-SR    &                           & 31.49 & 22.7  & 0.64 & 11.26 & 20.88 & 19.57 & 2.66 \\
\multirow{5}{*}{Diffusion}   & SR3       &                           & 27.7  & 23.75 & 0.64 & 13.42 & 18.7  & 18.52 & 3.2  \\
                             & IRSDE     &                           & 30.81 & 22.84 & 0.68 & 14.99 & 18.02 & 16.36 & 4.05 \\
                             & EDiffSR   &                           & 28.32 & 24.91 & 0.63 & 13.73 & 18.44 & 15.92 & 5.07 \\
                             & LWTDM     &                           & 29.35 & 19.05 & 0.65 & 12.78 & 18.09 & 19.78 & 3.25 \\
                             & Proposed  &                           & \textbf{21.04} & \textbf{27.76} & \textbf{0.87} & \textbf{9.71}  & \textbf{8.37}  & \textbf{15.12} & \textbf{7.89} \\
\multirow{2}{*}{GAN}         & LE-GAN    & \multirow{9}{*}{$\times 4$} & 33.62 & 19.4  & 0.62 & 12.69 & 12.69 & 19.77 & 2.04 \\
                             & ESRGAN-DP &                           & 32.97 & 18.65 & 0.59 & 18.38 & 16.39 & 23.13 & 3.44 \\
\multirow{2}{*}{VAE}         & RefVAE    &                           & 36.89 & 17.78 & 0.54 & 13.24 & 19.51 & 19.66 & 2.15 \\
                             & LSR-SR    &                           & 35.12 & 17.26 & 0.56 & 16.19 & 15.55 & 22.37 & 1.97 \\
\multirow{5}{*}{Diffusion}   & SR3       &                           & 32.23 & 18.73 & 0.58 & 15.61 & 14.62 & 19.11 & 2.59 \\
                             & IRSDE     &                           & 32.91 & 17.15 & 0.58 & 15.09 & 13.63 & 21.15 & 2.97 \\
                             & EDiffSR   &                           & 27.41 & 18.35 & 0.57 & 14.86 & 15.92 & 22.1  & 2.56 \\
                             & LWTDM     &                           & 31.8  & 18.63 & 0.53 & 14.32 & 15.17 & 24.39 & 1.79 \\
                             & Proposed  &                           & \textbf{24.63} & \textbf{27.71} & \textbf{0.82} & \textbf{10.04} & \textbf{9.71}  & \textbf{14.66} & \textbf{5.92} \\ \bottomrule
\end{tabular}
}
\end{table}

\begin{figure}[]   
    \centering  
    \includegraphics[width=5.5in]{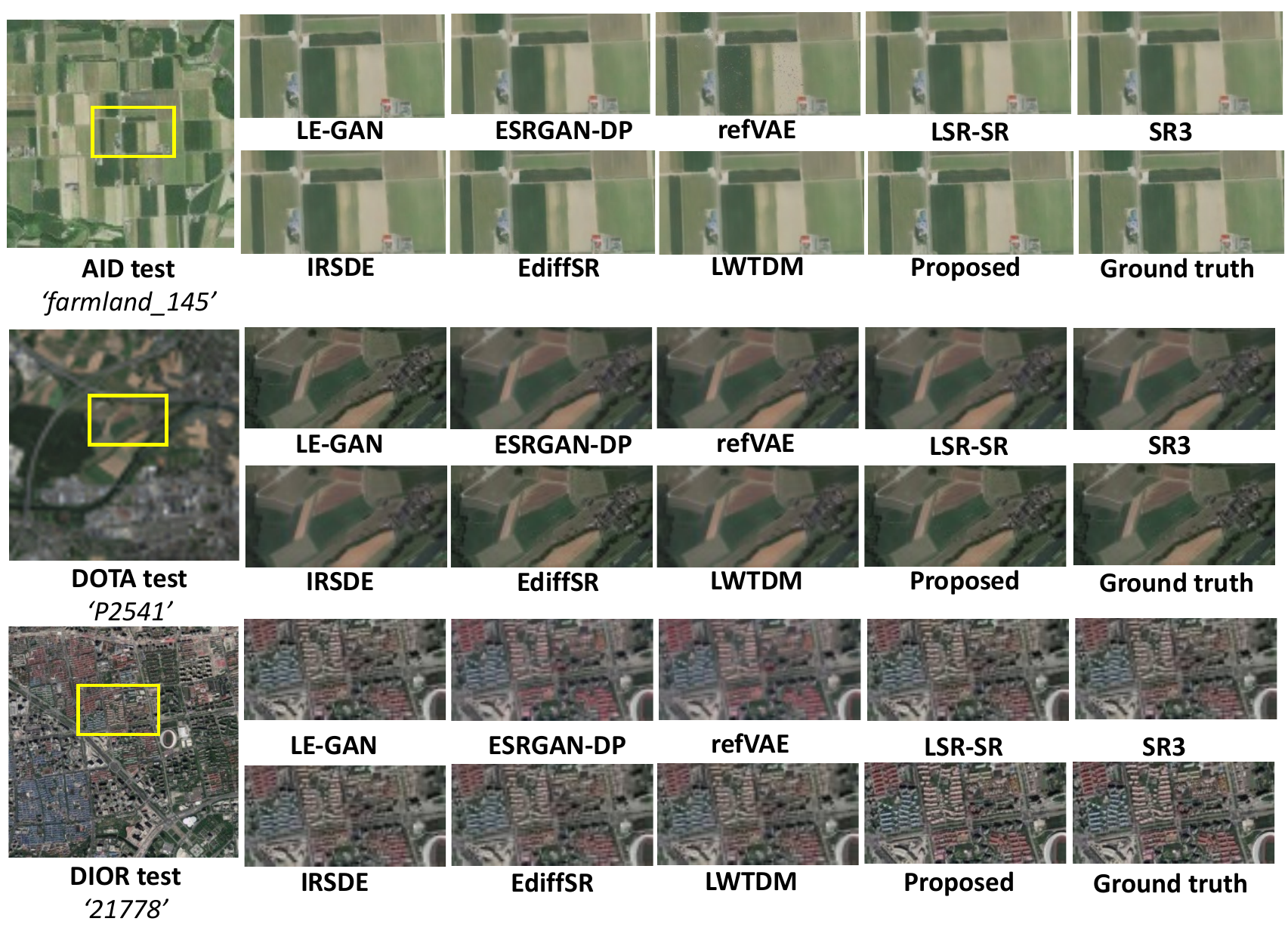}  
    \caption{ $\times 2$ visual comparisons with SOTA SR models on AID, DOTA, and DIOR test set. The square image at the left side is the input $64 \times 64$ image, and the rectangular frames indicate the zoomed-in view of the $\times 2$ super-resolution results for a better view.}
    \label{fig:3}  
\end{figure}

\textit{Qualitative Comparison.}  A visual comparison between the proposed model and the SOTA SR models is also conducted. Figure \ref{fig:3} demonstrates a comparisons of $\times 2$ super-resolution results ($64 \times 64 \rightarrow 128 \times 128$) on the AID,DOTA and DIOR test sets. Our proposed model consistently generates more realistic and detailed reconstruction closely resembling the ground truth image. In the case of the $"farmland_145"$ image from the AID test, the proposed model captures fine textures in farmland patches. the VAE-based models like refVAE and LWTDM show over-smoothed regions, losing critical texture details. In the case of the $"P2541"$ image from the DOTA test, the grass area reconstructed by the proposed model aligns more accurately with the ground truth, but competing models (e.g.refVAE and LSR-SR) exhibit blurred details, reflecting weaker spatial generalization. In the case of the $"21778"$ image from the DIOR test, the proposed model restores clear spatial details of densely built-up areas (as highlighted in the zoomed-in frames), outperforming other models like refVAE, LSR-SR, IRSDE, and LWTDM, which produce noticeable blurs. \par

\begin{figure}[]   
    \centering  
    \includegraphics[width=5.5in]{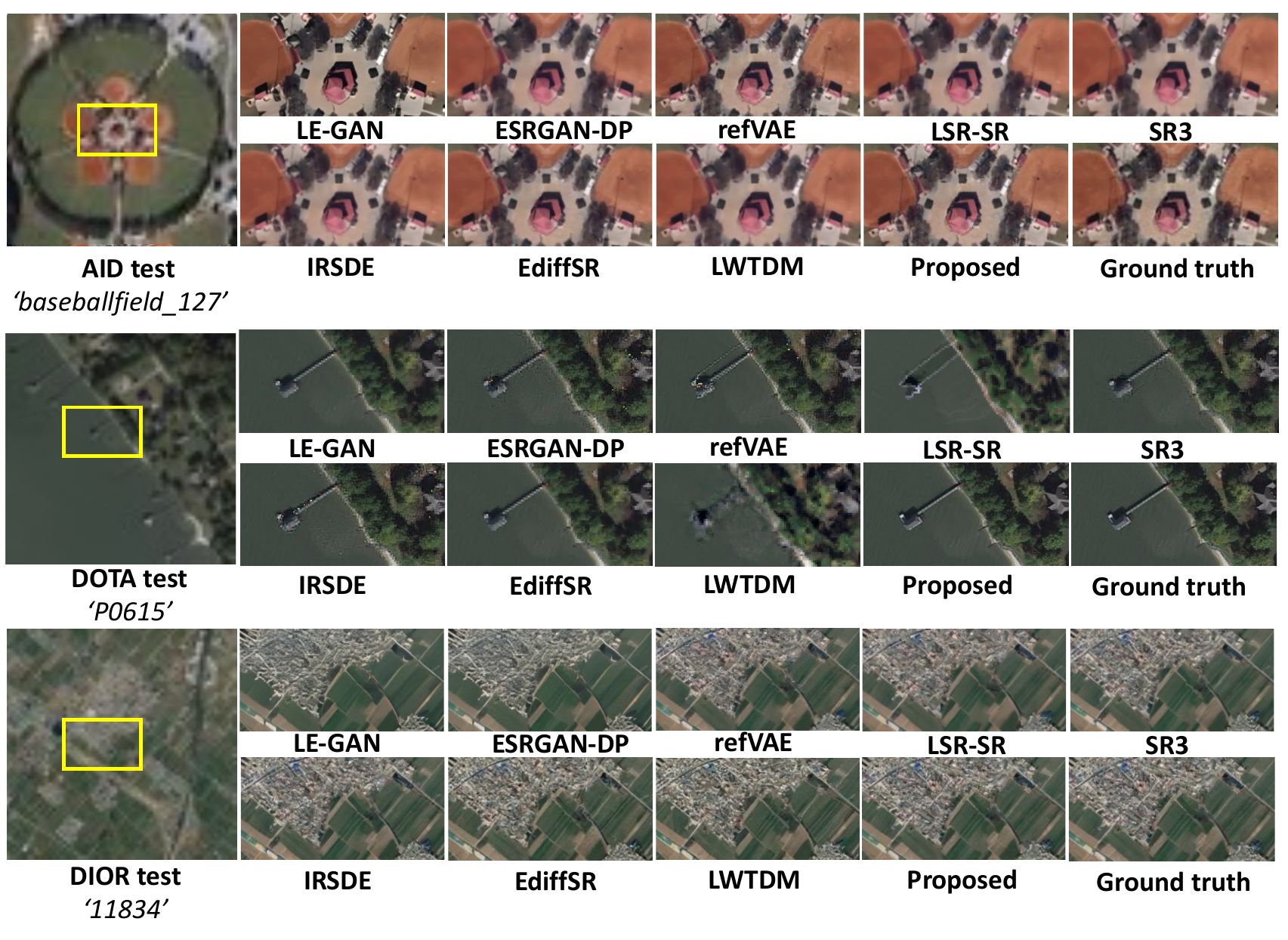}  
    \caption{$\times 4$ visual comparisons with SOTA SR models on AID, DOTA, and DIOR test set. The square image at the left side is the input $64 \times 64$ image, and the rectangular frames indicate the zoomed-in view of the $\times 4$ super-resolution results for a better view.}
    \label{fig:4}  
\end{figure}

Figure \ref{fig:4} provides a comparative analysis of visual results when increasing the super-resolution scale factor from $\times 2$ to $\times 4$ ($64 \times 64 \rightarrow 256 \times 256$). The analysis highlights the challenges and performance degradation by existing methods when increasing the scale factor. Most SOTA models exhibit noticeable degradation in performance. For example, blurring of fine structures in the $'baseballfield_127'$ image from the AID test, such as tree crowns, rood details and bridge, is prominent in VAE-based methods (e.g. RefVAE and LSR-SR) and diffusion-based methods (e.g. RSDE and LWTDM). This degradation is particularly pronounced in the DOTA and DIOR tests. More specifically, in the $'P0615'$ image from the DOTA test, the bridges reconstructed by RefVAE, LSR-SR, IRSDE, and LWTDM are plagued with textural blurring. In the $'11834'$ image from the DOTA test, LSR-SR and refVAE produce over-smoothed outputs, while other GAN-based and diffusion-based models provide more natural results with ground truth. However, GAN-based models like LE-GAN and ESRGAN-DP restore edge information but introduce noticeable color artefacts, diverging from the natural appearance of the ground truth. In contrast, our proposed model leverages the cross-scale pyramid architecture of the proposed model in predicting contextual prior to recover fine-grained texture, such as the details of bridge, enhancing the performance of diffusion models in super-resolution tasks.

\begin{figure}[]   
    \centering  
    \includegraphics[width=5.5in]{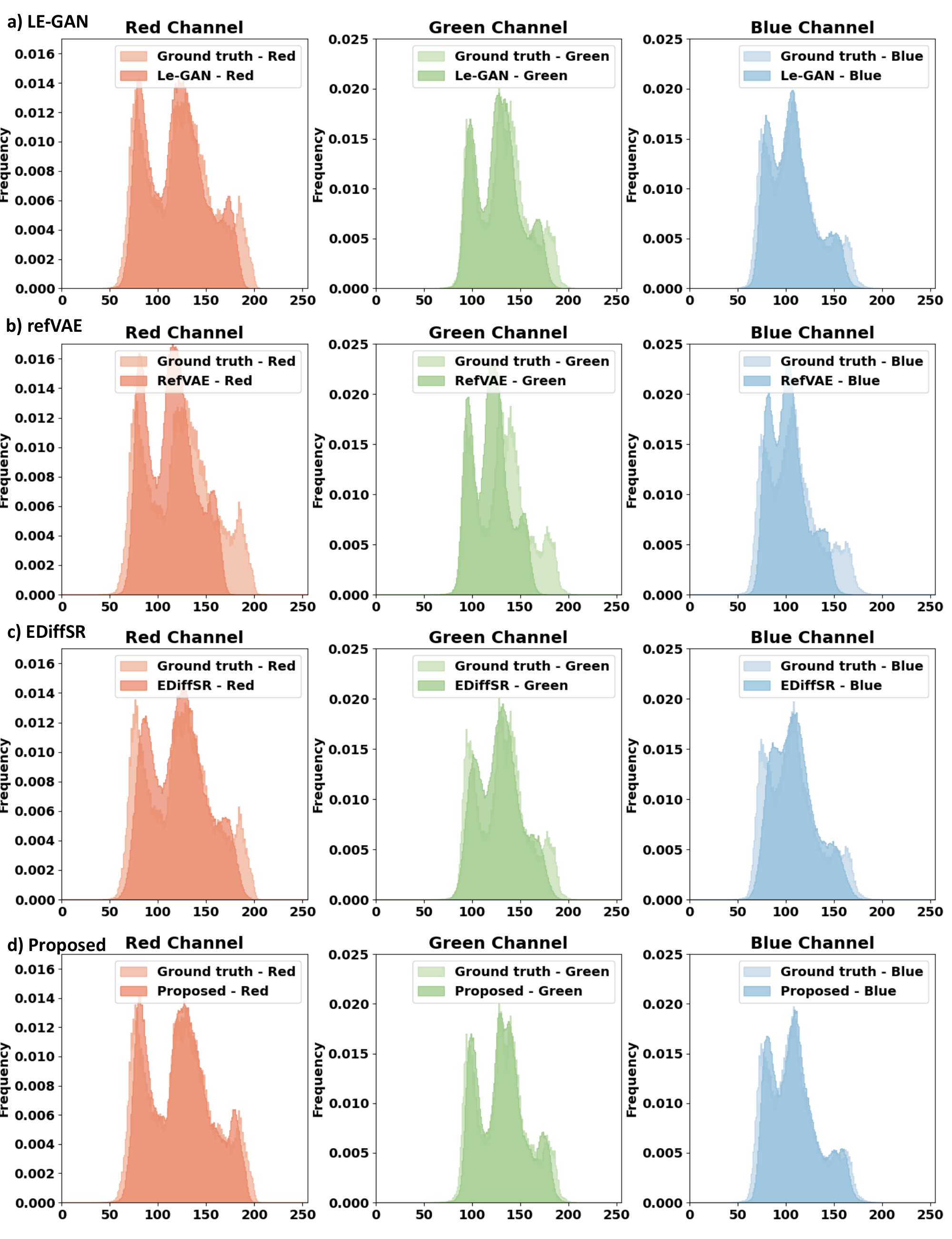}  
    \caption{A comparison of spectral distributions across the red, green, blue channels between the ground truth and the $\times 2$ super-resolution image produced by a) LE-GAN, b) refVAE, c)EDiffSR, and d) proposed model for the $'P0615'$ image from the DOTA test. All the values in red, green, and blue channels are normalized to the range of $[0, 255]$.}
    \label{fig:5}  
\end{figure}

\begin{figure}[]   
    \centering  
    \includegraphics[width=5.5in]{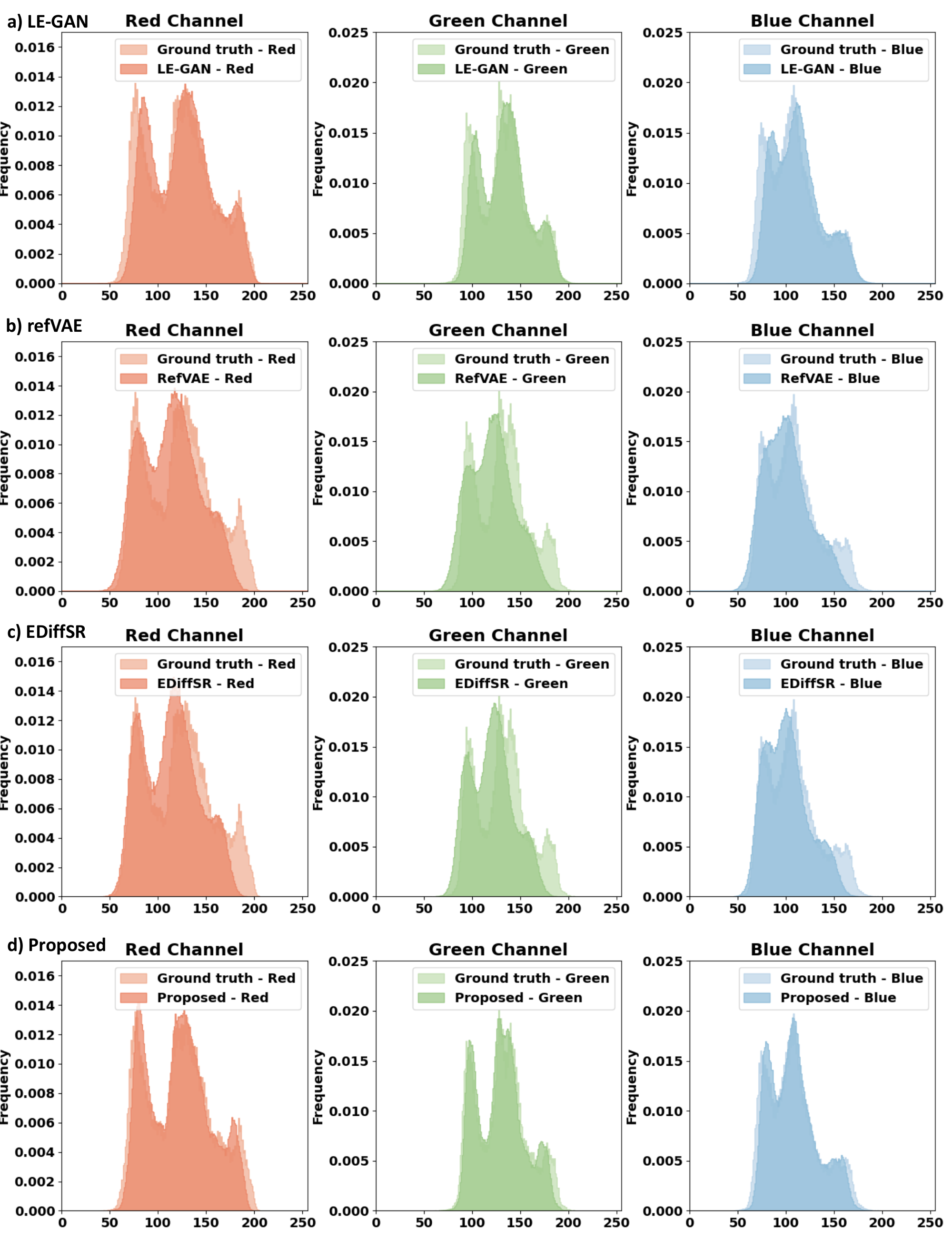}  
    \caption{A comparison of spectral distributions across the red, green, blue channels between the ground truth and the $\times 4$ super-resolution image produced by a) LE-GAN, b) refVAE, c)EDiffSR, and d) proposed model for the $'P0615'$ image from the DOTA test. All the values in red, green, and blue channels are normalized to the range of $[0, 255]$.}
    \label{fig:6}  
\end{figure}

\textit{Spectral channel evaluation.} Spectral consistency is crucial in evaluating the super-resolution performance for remote sensing images. Figure \ref{fig:5} and Figure \ref{fig:6} compare the distributions of color channels ( red, green and blue) for the $'P0615'$ image from the DOTA test at $\times 2$ and $\times 4$ scale factors. Representative models, LE-GAN, RefVAE, and EDiffSR, are selected as benchmarks for GAN-based, VAE-based, and diffusion-based methods, respectively. It can be observed that for the $\times 2$ super-resolution task all models can adequately reproduce the spectral distributions of the ground truth in all three channels, a characteristic double-peak pattern. However, when the scale factor increase to $\times 4$, the results, as shown in Figure \ref{fig:6}, are different. Only the proposed model maintains spectral consistency across the RFGB channels to the ground truth. The spectral distribution restored from the benchmarked models deviates from the the ground truth. This observation indicates the proposed model is more robust in restoring spectral details at larger scale factors.

% \begin{comment}

% Based on the preceding evaluation, this section focuses on using LE-GAN, RefVAE, and EDiffSR as representative models for GAN-based, VAE-based, and diffusion-based approaches, respectively. For $\times 2$ super-resolution, all candidate models achieve satisfactory likelihood estimation of the ground truth's spectral distribution, characterized by a double-peak pattern. Notably, both LE-GAN and our proposed model successfully estimate the spectral distribution across the red, green, and blue channels. However, the scenario changes for $\times 4$ super-resolution (i.e. Figure \ref{fig:6}), where only the proposed model accurately restores the spectral details consistent with the ground truth image. The RGB channel restored from LE-GAN, refVAE and EDiffSR deviate from the ground-truth distribution. These findings illustrate that the proposed model is robust in restoring spectral details for large scale factors, and enhanced the performance of diffusion models in super-resolution tasks.
% \end{comment}

\subsection{Evaluation for UR tasks}
\label{sec:441}
In this section, we present the evaluation results of the proposed WaveDiffUR SDE solver for UR tasks with a $\times 2$ scale factor. Again, we used LE-GAN, refVAE, and EDiffSR as representative models from the GAN, VAE and diffusion families,respectively, serving as the upscaling modules in the WaveDiffUR framework. To further assess the effectiveness of the proposed self-cascade UR strategy, we benchmarked WaveDiffUR against a one-step UR strategy, which directly fine-tuned SR models for UR tasks. All models were tested on the DOTA and DIOR testing sets for simulated evaluation and on the DDISR dataset for real-world evaluation.\par

In the simulated evaluation, the DOTA and DIOR testing sets, where the highest-resolution data was treated as ground truth, and simulated lower-resolution data was generated using bicubic down-sampling. For UR resolutions beyond the highest available ground truth, pseudo ground truth images were generated by upsampling the ground truth to match the UR results for evaluation. PNSR and SRE were employed to evaluate the spatial fidelity and spectral consistency, respectively. The results are presented in Table. \ref{tab:4}. The proposed CSP-WaveDiffUR model achieves the highest PSNR and SRE scores across all independent datasets and for all upscaling factors evaluated. It significantly outperforms its competitors, with an approximate 1.5 times improvement in PSNR and SRE for large-scale ($> \times 32 $) UR. At extreme magnifications (e.g., $\times 128$), it achieves up to approximate 3 times improvement in PSNR and SRE. This is due to the Cross-Scale Pyramid (CSP) constraint in the proposed model which effectively guides the diffusion-denoising inference, enabling accurate reconstruction of both spectral and spatial details in UR tasks. Moreover, the results indicate that SR pipelines (LE-GAN, refVAE and EDDiffSR) embedded in the WaveDiffUR architecture outperform their one-step fine-tune counterparts, highlighting the benefit of the self-cascade approach. By iteratively refining UR images, the model ensures realistic inferences.

\begin{table}[]

\caption{A quantitative comparison of model performance in terms of PSNR and SRE metrics when different SR backbone models are integrated into the WaveDiffUS SDE solver for the UR tasks (scaling from $\times 4$ to $\times 128$) using the independent datasets DOTA, DIOR, and DDISR. Outputs marked with an asterisk (*) indicate evaluations are performed with pseudo ground truth images, where the UR image resolution exceeds the ground truth resolution.}
\label{tab:4}

\centering
\resizebox{6in}{!}{
\begin{tabular}{ccccccc}
\toprule
                                             & \multicolumn{6}{c}{DOTA}                                                          \\\midrule
 \multicolumn{1}{c}{\multirow{2}{*}{Method}} & \multicolumn{6}{c}{Scale}                                                         \\
 \multicolumn{1}{c}{}                        & $\times 4$  & $\times 8$  &$\times 16$  &$\times 32$* &$\times 64$* & $\times 128$* \\\midrule
 LE-GAN                                     & 19.15/11.41 & 17.11/12.88 & 15.91/13.32 & 11.53/15.42 & 9.21/18.89  & 7.91/23.33  \\
 LE-GAN-WaveDiffUR                           & 19.74/11.39 & 19.18/11.12 & 18.21/12.86 & 16.74/13.40 & 12.58/13.63  & 11.21/15.87  \\
 RefVAE                                      & 22.14/14.22 & 22.01/15.98 & 20.34/16.23 & 16.34/18.23 & 12.01/23.99 & 10.34/30.24 \\
 RefVAE-WaveDiffUR                            & 22.38/13.13 & 22.82/14.25 & 21.38/14.14 & 19.38/15.14 & 18.32/16.26 & 17.38/19.15 \\
 EDiffSR                                     & 20.15/16.31 & 18.92/17.31 & 15.15/18.32 & 10.15/22.32 & 8.92/27.32  & 7.15/38.33  \\
 EDiffSR-WaveDiffUR                          & 20.02/16.51 & 19.29/17.02 & 18.02/17.52 & 17.25/18.21 & 15.99/18.35  & 12.02/20.53  \\
 CSP-WaveDiffUR(one-step)                    & 28.21/7.82  & 28.18/7.88  & 26.21/9.83  & 21.21/12.83 & 19.18/17.89 & 18.21/19.84 \\
 CSP-WaveDiffUR                              & 28.32/7.22  & 28.31/7.37  & 27.32/8.03  & 25.52/9.23 & 24.31/11.38 & 24.28/13.24 \\\midrule
                                             & \multicolumn{6}{c}{DIOR}                                                          \\\midrule
 \multicolumn{1}{c}{\multirow{2}{*}{Method}} & \multicolumn{6}{c}{Scale}                                                         \\
 \multicolumn{1}{c}{}                        &$\times 4$  & $\times 8$  &$\times 16$*  &$\times 32$* &$\times 64$* & $\times 128$* \\\midrule
 LE-GAN                                   & 19.41/12.69 & 19.19/13.18 & 18.31/14.82 & 14.35/17.34 & 10.21/19.99 & 6.92/27.53  \\
 LE-GAN-WaveDiffUR                        & 19.98/12.46 & 19.33/13.08 & 18.42/14.66 & 17.21/16.89 & 15.99/17.72 & 14.25/20.55  \\
 RefVAE                                   & 18.35/15.92 & 17.32/16.18 & 15.52/16.74 & 12.21/19.98 & 8.81/25.93  & 5.89/28.83  \\
 RefVAE-WaveDiffUR                        & 19.01/15.28 & 18.99/15.82 & 17.59/16.11 & 14.49/17.52 & 13.21/19.14 & 10.15/21.71  \\
 EDiffSR                                  & 18.78/13.51 & 17.98/14.23 & 14.88/15.66 & 12.62/20.22 & 9.21/24.41  & 5.32/26.93  \\
 EDiffSR-WaveDiffUR                       & 18.11/12.82 & 17.23/13.31 & 16.32/14.83 & 15.12/16.11 & 13.11/18.21  & 10.31/19.68  \\
 CSP-WaveDiffUR(one-step)                 & 24.63/12.71  & 23.21/13.81  & 21.92/15.32 & 18.84/19.32 & 14.22/22.12 & 10.36/24.96 \\
 CSP-WaveDiffUR                           & 25.12/8.99  & 24.92/9.87  & 22.31/9.72  & 21.03/10.21 & 18.79/12.42 & 16.32/14.85 \\\midrule
                                             & \multicolumn{6}{c}{DDISR}                                                         \\\midrule
 \multicolumn{1}{c}{\multirow{2}{*}{Method}} & \multicolumn{6}{c}{Scale}                                                         \\
 \multicolumn{1}{c}{}                        & $\times 4$  & $\times 8$  &$\times 16$  &$\times 32$  &$\times 64$  & $\times 128$ \\\midrule
LE-GAN                                     & 24.14/16.89 & 21.91/18.87 & 19.19/21.82 & 16.59/24.34 & 12.21/28.99 & 8.21/32.31  \\
LE-GAN-WaveDiffUR                          & 24.38/15.08 & 23.21/16.16 & 23.09/16.98 & 19.99/19.72 & 16.55/22.55 & 14.32/26.37 \\
RefVAE                                     & 28.12/18.18 & 25.52/19.74 & 22.11/21.81 & 18.81/25.93 & 12.89/28.23 & 9.82/36.20  \\
RefVAE-WaveDiffUR                          & 28.29/17.21 & 27.59/18.91 & 26.39/19.12 & 22.99/20.41 & 19.15/23.71 & 16.15/27.71 \\
EDiffSR                                     & 28.98/23.13 & 27.88/26.66 & 24.92/29.12 & 19.21/34.19  & 13.28/39.93  & 12.08/38.82  \\
EDiffSR-WaveDiffUR                          & 29.23/21.11 & 28.32/22.83 & 27.17/23.11 & 25.91/24.29  & 21.18/25.68  & 18.11/28.85  \\
CSP-WaveDiffUR(one-step)                    & 31.21/12.19  & 30.92/13.39 & 28.84/16.32 & 24.22/18.12 & 20.36/21.62 & 16.13/28.21 \\
CSP-WaveDiffUR                              & 34.92/8.27  & 34.19/8.75  & 32.33/9.26  & 30.79/11.95 & 28.32/14.85 & 23.48/16.85 \\ \bottomrule
\end{tabular}
}
\end{table}

\begin{figure}[]   
    \centering  
    \includegraphics[width=5.5in]{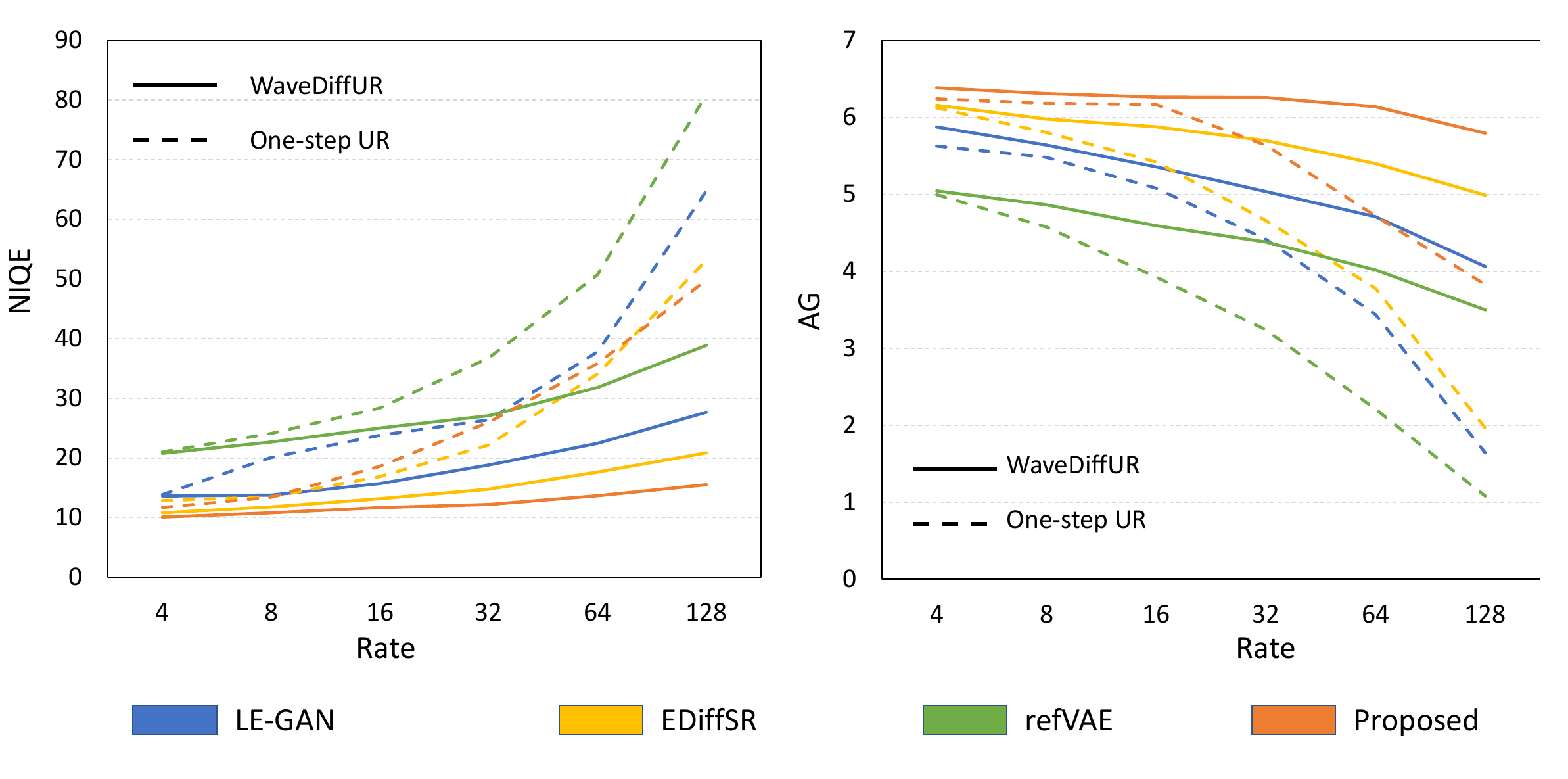}  
    \caption{A comparison of model performance between WaveDiffUR-based models with different backbone models (i.e. LE-GAN, EDiffSR and RefVAE), the traditional one-step fine-tune method and the proposed CSP-WaveDiffUR model for UR tasks in terms of NIOE and AG  on large-scale UR in terms of NIQE and AG.}
    \label{fig:7}  
\end{figure}

To validate the real-world applicability and generalization, we evaluated the proposed model on the real-world DDISR dataset without applying simulated degradations. As presented in Table. \ref{tab:4}, the proposed CSP-WavediffUR model achieves the highest PSNR and SRE scores across all upscaling factors. It significantly outperforms its competitors, achieving an approximately 1.6 times improvement in PSNR and 1.8 times reduction in SRE for large-scale UR (magnifications $> \times 32$). At extreme magnifications (e.g., $\times 128$), it achieves up to a 2.8 times improvement in PSNR and SRE. Figure \ref{fig:7} illustrates a quantitative comparison of model performance between WaveDiffUR-based models with different backbone models (i.e. LE-GAN, EDiffSR and RefVAE), the traditional one-step fine-tune method and the proposed CSP-WaveDiffUR model for UR tasks in terms of NIOE and AG metrics. The performance degradation is observed across all models in large-scale UR tasks, but the proposed WaveDiffUR-based models with different backbone models demonstrate the most robust performance, approximately with 2.2 times better in NIQE and AG for $\times 128$ magnificant scale. The enhanced approach, CSP-WaveDiffUR, achieves the best performance in NIQE and AG (i.e. $11.8\%$ and $19.1\%$ performance degradation in NIQE and AG for mangificant sacle growth from $\times 4$ to $\times 128$, respectively), indicating that the proposed method can reconstruct images that contain more high-frequency detail information and align well with human perception in real-world scenarios.\par

% \begin{comment}

% To further demonstrate the generalization capabilities of UR models in real-world applications, we also evaluate the performance of the proposed model on the real-world DDISR dataset without performing simulated degradations. Figure \ref{fig:7} presents the quantitative comparison between the proposed model with different SR backbone models and the traditional one-step fine-tune approach for UR task in terms of non-referenced metrics, NIQE and AG. The performance degradation is observed across all models in large-scale UR tasks, but the proposed WaveDiffUR-based models with different backbone models demonstrate the most robust performance. Furthermore, the enhanced approach, CSP-WaveDiffUR, achieves the best performance in NIQE and AG, indicating that the proposed method can reconstruct images that contain more high-frequency detail information and align well with human perception in real-world scenarios.\par
% \end{comment}

\begin{figure}[]   
    \centering  
    \includegraphics[width=5.5in]{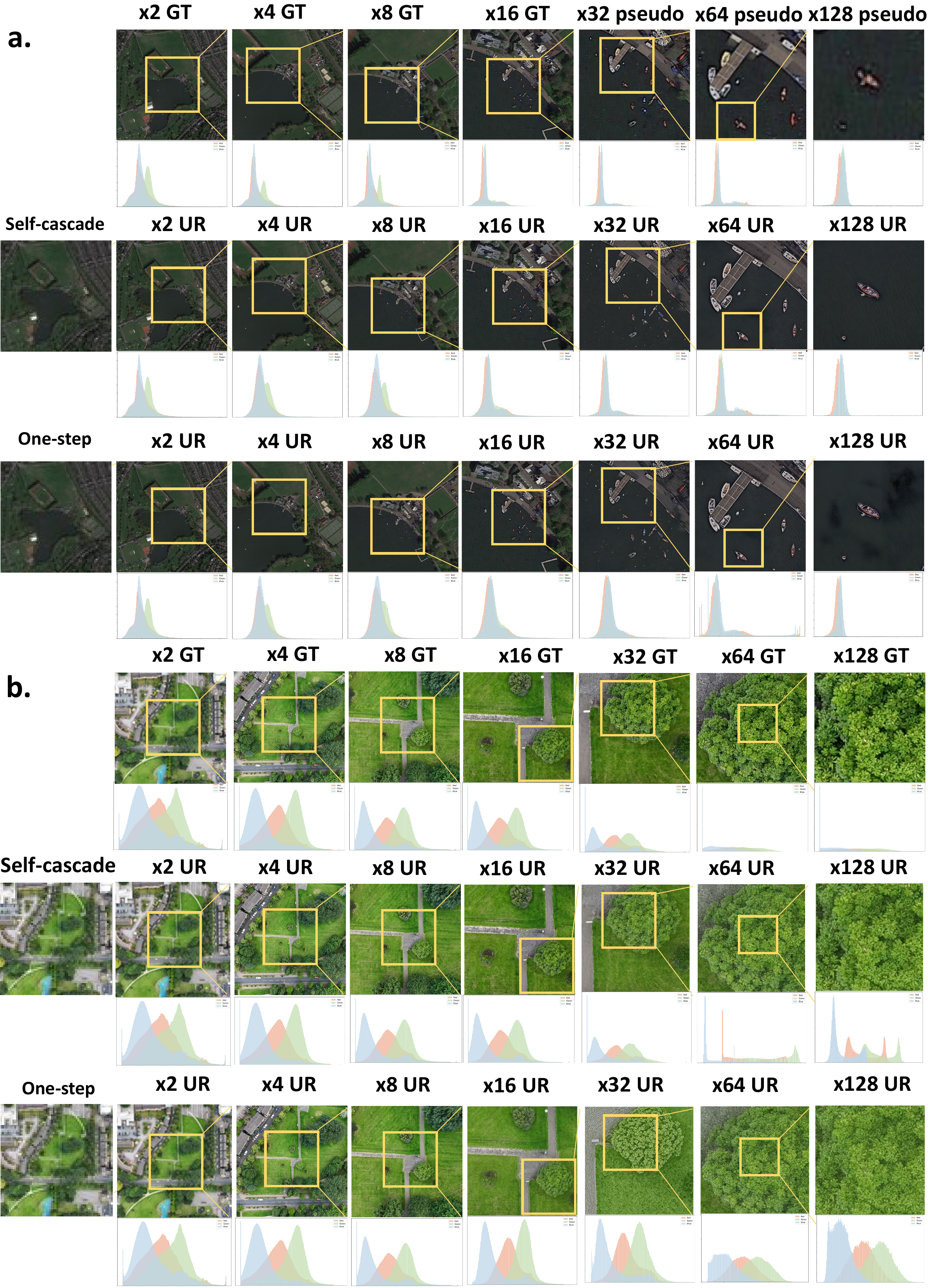}  
    \caption{Visual comparisons of UR images generated from the one-step and self-cascade approaches in terms of spatial fidelity and channel consistency in a) simulated evaluation and b) real-world evaluation. For the simulated evaluation, the ground truth image is generated by zoom-in or zoom-out of the raw HR image. For the real-world evaluation, the ground truth image is collected by the multi-scale measurements. Zoom-in view in the yellow frame for a better view.}
    \label{fig:8}  
\end{figure}

To provide an intuitive comparison, we present a visual analysis of simulated and real-world evaluations. The qualitative results of the CSP-WaveDiffUR model, utilizing one-step and self-cascade strategies, are shown in Figure \ref{fig:8}. We observe that when the UR rate is below $\times 32$, both the one-step and self-cascade approaches perform similarly, effectively recovering the overall spatial details. However, as the UR rate exceeds $\times 32$, the one-step approach begins to exhibit a blurring effect, with noticeable degradation in spatial fidelity and channel consistency, especially when compared to the self-cascade method. In the comparison of UR images with ground truth HR images at higher UR rates (such as $\times 64$ and $\times 128$), the UR images can still recover accurate details for simpler objects (e.g., white lines on the bay and embankment in simulated evaluations, or roads in real-world evaluations). However, they tend to introduce exaggerated sharpening with pseudo-details for more complex structures (e.g., boats in simulated evaluations or tree canopies in real-world evaluations). This issue is particularly noticeable with the one-step approach. A potential explanation is the accumulation of cross-scale biases during the transition, which hampers the realistic reconstruction of high-frequency components, leading to overly sharp, unnatural details.\par

\subsection{Ablation Studies}
In this section, we present extensive experiments to demonstrate the effectiveness of each component within our self-cascade model.\par

1) \textit{Component analysis}: To evaluate the effectiveness of each component in the proposed self-cascade model, we conducted an ablation study by removing three key elements: CSP conditions, high-frequency restoration, and self-cascade structure one by one. This produced three simplified models, Baseline, Model-1 and Model-2. Table \ref{tab:5} summarises the model configurations and the performance evaluation results in terms of FID, SRE and NIQE. We set the upscale factor to $\times 8$ as an example. When the self-cascade architecture was excluded, the model was trained directly for the $\times 8$ upscale factor. As shown in Table \ref{tab:5}, the evaluation results of  Model-1 (with CSP condition) and the baseline diffusion model indicate that the CSP constraint significantly improves spatial fidelity (FID: 63.62 versus 81.28) while reducing model size (98.24M versus 168.14M). The evaluation results of Model-2 (with high-frequency restoration) and the baseline model display that introducing the high-frequency restoration improves channel consistency (SRE: 19.15 vs. 27.74). When both CSP constraints and the high-frequency restoration module are combined into the Proposed-1 model, a slight increase in model parameters is observed, but the FID and SRE metrics improve substantially.
Embedding the Proposed-1 model into the self-cascade architecture requires no additional external parameters, as the same model is reused without fine-tuning. This integration yields significant improvements across all metrics, including FID, SRE, and NIQE. These findings demonstrate the effectiveness of the proposed methodology in enhancing large-scale image upscaling. Moreover, the low-complexity design of the components ensures that the self-cascade architecture remains both efficient and powerful.

\begin{table}[]
\caption{Ablation analysis of the proposed methodology in the $\times 8$ UR task, and the progressive self-cascade strategy is used as an example.}
\label{tab:5}
\centering
\resizebox{6in}{!}{
\begin{tabular}{cccccccc}
\toprule
Model      & $\tau_{\theta}$ & High-frequency & Self-cascade   & Param.(M) & FID   & SRE   & NIQE  \\\midrule
Baseline   &                 &                &         & 36.84    & 81.28 & 27.74 & 18.89 \\
Model-1    & $\checkmark$    &                &         & 24.82     & 63.62 & 26.49 & 16.82 \\
Model-2    &                 & $\checkmark$   &         & 14.14    & 87.82 & 19.15 & 15.52 \\
Proposed-1 & $\checkmark$    & $\checkmark$   &         & 29.12    & 59.15 & 16.24 & 14.33 \\
Proposed-2 & $\checkmark$    & $\checkmark$   & $\checkmark$   & 29.12    & 54.82 & 9.81  & 14.03 \\\bottomrule
\end{tabular}
}
\end{table}

\begin{figure}[]   
    \centering  
    \includegraphics[width=4.5in]{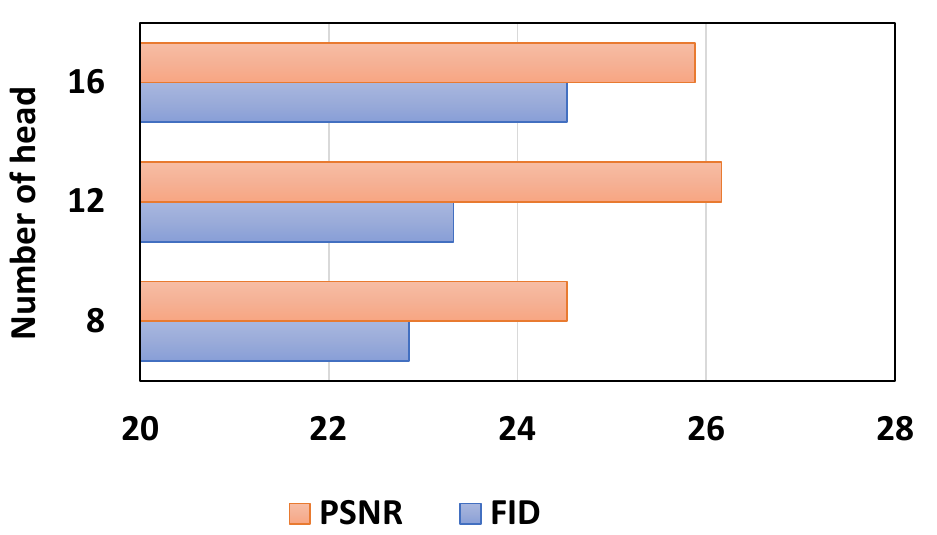}  
    \caption{Ablation analysis of CSP constraint conditions with different heads of cross-attention blocks in terms of FID and SRE.}
    \label{fig:9}  
\end{figure}

2) \textit{Effectiveness of CSP constraints}: We analyzed the impact of varying the number of heads in the cross-attention blocks for modeling CSP constraint conditions. As shown in Figure \ref{fig:9}, the proposed model achieves slightly better FID performance with 16 heads compared to 12 heads, while the highest PSNR results are obtained with 12 heads. To balance model size and performance effectively, we set the default number of heads to 12.

The impact of the number of heads within the self-cascade architecture on the model performance was also conducted, 8, 12 and 16 heads were analysed in terms of FID and SRE. The experimental results, as shown in Table \ref{tab:6}, show that the default configuration (12 heads) delivers a balanced trade-off on UR performance, optimizing both FID and SRE metrics.
\par

\begin{table}[]

\caption{Ablation analysis of the number of heads of the cross-attention blocks in self-cascade architecture.}
\label{tab:6}
\centering
\resizebox{4in}{!}{
\begin{tabular}{cccc}
\toprule
            & \multicolumn{3}{c}{Number of head}            \\\midrule
Models      & 8           & 12               & 16           \\
Reuse       & 24.47/10.82 & 24.21/9.81       & 23.78/10.12  \\
Progressive & 24.12/8.99  & 23.92/8.87       & 23.71/8.91   \\ \bottomrule
\end{tabular}
}
\end{table}

\begin{figure}[]   
    \centering  
    \includegraphics[width=4.5in]{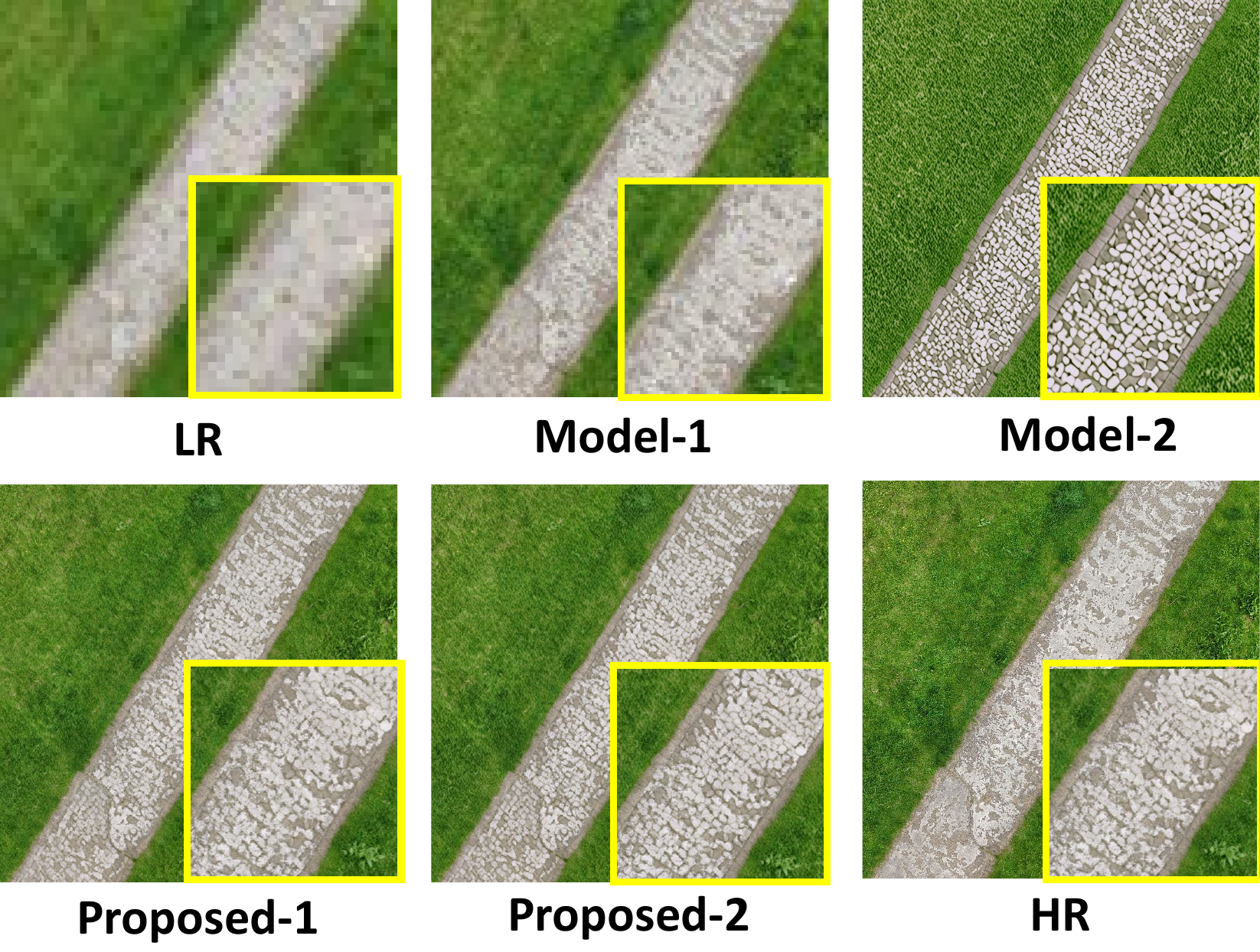}  
    \caption{Visual ablation analysis of the $\times 128$ UR results from the model configurations listed in Table \ref{tab:5}.}
    \label{fig:10}  
\end{figure}

3) \textit{Effectiveness of high-frequency restoration}: To demonstrate the capability of high-frequency restoration in recovering fine details for accurate UR reconstruction, we present a visual comparison in Figure \ref{fig:10}. 
Comparing the UR outputs of Model-1 (cross-scale condition-only) and Model-2 (high-frequency predictor-only), it is evident that Model-2 produces overly sharpened details compared to Model-1. Comparing the UR results of Proposed-1 and Proposed-2 within the self-cascade architecture, the combination of the high-frequency predictor and cross-scale condition achieves a superior balance between high-frequency and low-frequency components. This is particularly noticeable in features such as road edges and surface textures. These observations emphasize that the high-frequency predictor significantly enhances UR performance by predicting enriched high-frequency details with more priors, effectively improving overall reconstruction quality.

\begin{figure}[]   
    \centering  
    \includegraphics[width=4.5in]{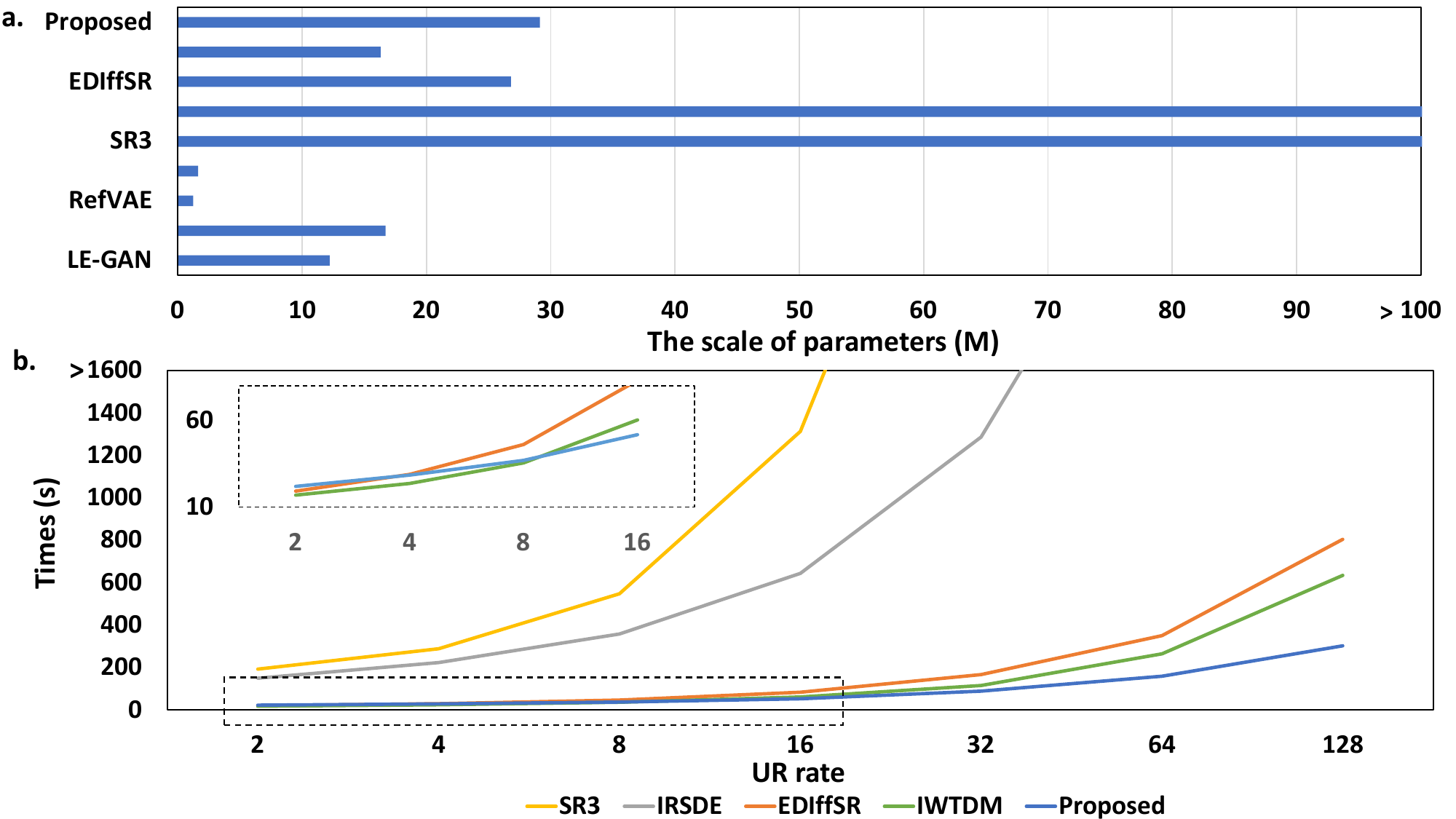}  
    \caption{Model efficient comparison regarding a) the scale of baseline parameters, and b) the self-cascade model execution times with various UR rates.}
    \label{fig:11}  
\end{figure}

4) \textit{Model efficiency}: To evaluate the efficiency of the self-cascade strategy, we compared the parameters and execution times of the baseline model with those of the self-cascade architecture. As shown in Figure \ref{fig:11}, while the proposed baseline model is not the lightest among existing DPM-based SR models, it remains highly competitive in terms of efficiency.

Notably, the proposed model within the self-cascade UR architecture demonstrates faster inference speeds compared to its competitors, especially for large-scale UR tasks (>$\times 16$). This efficiency gain is primarily due to the wavelet transformation integrated into our model. By compressing the input data and processing high-frequency and low-frequency components in parallel at each UR scale, the wavelet transformation leads to reduced memory requirements for storing intermediate feature maps and fewer convolutional operations, thereby lowering computational costs. These optimizations enable the self-cascade architecture to outperform other approaches in computational efficiency, making our method more practical and scalable for real-world applications.

\section{Conclusions}
\label{sec:con}

In this study, we introduce the WaveDiffUR architecture, designed to address the ill-posed challenges inherent in remote sensing super-resolution (SR) and ultra-resolution (UR) tasks. This architecture generates spectral-spatial consistent and perceptually pleasing SR and UR results for remote sensing images. Building on this foundation, the proposed CSP-WaveDiffUR model demonstrates superior performance compared to state-of-the-art (SOTA) methods in solving the UR stochastic differential equation (SDE).
Unlike traditional approaches that rely on interpolated conditions, our method incorporates CSP constraint conditions, which leverage priors derived from cross-scale spectral-spatial unmixing rules. This innovation effectively mitigates the ill-posed nature of the SDE solution, resulting in mitigating degradation across quantitative accuracy, perceptual quality, spectral consistency, and sharpness detail. Specifically, for quantitative accuracy, CSP-WaveDiffUR achieves approximately 3 times the improvement in PSNR compared to benchmark models at extreme magnifications (e.g., $\times 128$). In terms of perceptual quality and sharpness, the model demonstrates the most robust performance, with only $11.8\%$ and $19.1\%$ degradation in NIQE and AG metrics, respectively, as the magnification scale increases from $\times 4$ to $\times 128$, ranking first among benchmark evaluations. For spectral consistency, CSP-WaveDiffUR achieves up to 2 times reduction in SRE at $\times 128$ magnifications. Qualitative results further highlight the model’s ability to preserve fine-grained details and deliver superior visual fidelity, outperforming GAN-based, VAE-based, and diffusion-based SR models.\par

Despite its advantages, the proposed method has certain limitations. First, the modeling of CSP constraints requires high-quality low-resolution (LR) and reference (Ref) synchronous observation pairs, which restricts its applicability in regions with sparse observational data. Second, the method does not fully account for degradation variability across different remote sensing systems, reducing its adaptability to real-world scenarios.\par

To address these challenges, future research will prioritize two key areas: (1) Reducing dependency on reference images by developing techniques to tackle blind super-resolution (SR) problems, thereby broadening the method’s applicability in regions with limited or incomplete data; and (2) Enhancing robustness to degradation variability by designing mechanisms capable of modeling and adapting to diverse degradation patterns across different remote sensing platforms. These advancements aim to improve the robustness, adaptability, and scalability of the WaveDiffUR framework. With its demonstrated ability to outperform existing methods in accuracy and computational efficiency, the proposed approach has the potential to revolutionize applications in environmental monitoring, urban planning, disaster response, and precision agriculture, providing actionable insights at unprecedented resolutions. \par
\section*{Acknowledgments}
This work is supported by BBSRC.(BB/Y513763/1), EPSRC (EP/X013707/1). 

%Bibliography
\bibliographystyle{unsrt}  
\bibliography{references}

\end{document}